\documentclass[preprint,authoryear,12pt]{elsarticle}

\usepackage{epsfig}
\usepackage{amssymb}
\usepackage[ps2pdf,%
a4paper=true,%
breaklinks=true,%
colorlinks=true,%
pdfauthor={Smith et al.},%
pdftitle={The X-ray emission of the $\gamma$ Cassiopeiae stars}%
]{hyperref}

\newcommand{\gc}{$\gamma$\,Cas}

\newcommand{\xmmn}{XMM-Newton}
\newcommand{\xmm}{XMM}

\def\gtrsim{\mathrel{\hbox{\rlap{\hbox{\lower4pt\hbox{$\sim$}}}\hbox{$>$}}}}
\def\ltsim{\mathrel{\hbox{\rlap{\hbox{\lower4pt\hbox{$\sim$}}}\hbox{$<$}}}}

\journal{Advances in Space Research}

\begin{document}

%%%%%%%%%%%%%%%%%%%%%%%%%%%%%%%%%%%%%%%%%%%%%%%%%%%%%%%%%%%%%%%%%%%%%%%%%%%%%
%% Frontmatter
\begin{frontmatter}


% Use the tnoteref command within \title and fnref within \author or \address for footnotes;
% use the corref command within \author for corresponding author footnotes;
\title{The X-ray emission of the $\gamma$ Cassiopeiae stars}
% \author{Name\corref{cor1}\fnref{label2}}
% \address{Address\fnref{label3}}

% \title{Advances in Space Research Review on $\gamma$ Cassiopeiae}\tnotetext[footnote1]{This template can be used for all publications in Advances in Space Research.}

% Use optional labels to link authors explicitly to addresses:
% \author[Smith,Lopes de Oliveira, Motch]{}
%\address[NOAO]{}
%\address[Brazil]{}
% \address[France]{}

\author{Myron A. Smith}
%\corref{cor} }
%\fnref{footnote2}}
\address{National Optical Astronomical Observatory, 950 N. Cherry Ave., 
Tucson, AZ, U.S.A.}
%\cortext[cor]{Corresponding author}
%\fntext[footnote2]{Additional information regarding the corresponding author}
\ead{masmith@noao.edu}

% Url can be given like this:
% \ead[url]{http://www.elsevier.com/wps/find/authorsview.authors/latex}

\author{R. Lopes de Oliveira}
%\fnref{footnote3}}
\address{Universidade Federal de Sergipe, Departamento de F\'isica,
Av. Marechal Rondon, S/N, 49000-000 S\~ao Crist\'ov\~ao, SE, Brazil;
Observat\'orio Nacional, Rua Gal. Jos\'e Cristino 77, 20921-400, Rio de
Janeiro, RJ, Brazil }
\ead{rlopes@ufs.br/}

\author{C. Motch}
%\fnref{footnote4}}
\address{Observatoire Astronomique,  Universit\'e de Strasbourg, CNRS UMR7550, 
11 rue de  l'Universit\'e, F-67000, Strasbourg, France}
%\fntext[footnote4]{Additional information about the coauthors}
\ead{christian.motch@unistra.fr}

\begin{abstract}
Long considered as the ``odd man out" among X-ray emitting Be stars,
\gc\ (B0.5e\,IV) is now recognized as the prototype of a class of stars that
emit hard thermal X-rays.  Our classification differs from the historical 
use of the term ``\gc\ stars" defined from optical properties alone.
The luminosity output of this class 
contributes significantly to the hard X-ray production in massive stars in the
Galaxy.  The \gc\ stars have light curves showing variability 
on a few broadly-defined timescales and spectra indicative of an optically 
thin plasma consisting of one or more hot thermal components. 
By now  9--13 Galactic $\approx$B0-1.5e main sequence stars are 
judged to be members or candidate members of the \gc\ class. 
Conservative criteria for this designation
are for a $\approx$B0-1.5e III-V star to have an X-ray luminosity of
10$^{32}$--10$^{33}$ ergs\,s$^{-1}$, a hot thermal spectrum containing the
short wavelength  Ly$\alpha$ Fe\,XXV and Fe\,XXVI lines and the 
fluorescence FeK feature all in emission. 
If thermality cannot be demonstrated, for example from either the
presence of these Ly$\alpha$ lines or curvature of the hard continuum
of the spectrum of an X-ray active Be star, we call them
\gc\ {\it candidates.} 
We discuss the history of the discovery of the complicated 
characteristics of the variability in the optical, UV, and X-ray domains, 
leading to suggestions for the physical cause of the production of hard 
X-rays. These include scenarios in which matter from the Be star accretes 
onto a degenerate secondary star and interactions between magnetic fields 
on the Be star and its decretion disk.
The greatest aid to the choice of the causal mechanism is
the temporal correlations of X-ray light curves and spectra with
diagnostics in the optical and UV wavebands. We show why the 
magnetic star-disk interaction scenario is the most tenable 
explanation for the creation of hard X-rays on these stars.

\end{abstract}


\end{frontmatter}

\parindent=0.5 cm

%%%%%%%%%%%%%%%%%%%%%%%%%%%%%%%%%%%%%%%%%%%%%%%%%%%%%%%%%%%%%%%%%%%%%%%%%%%%%
%% Main text

\section{Introduction: description of \gc~ as an X-ray emitter.}

\subsection{Early X-ray discoveries.}

% General description -

Discovered as the first star to show Balmer line emission in
its spectrum \citep{Secchi}, \gc~ has long been held 
out as {\it the prototype} for what became known by the mid-20th century 
as a large class of ``classical Be" variables.\footnote{Classical
Be stars are main sequence or giant B stars whose spectra have 
shown Balmer line emission and have shown no evidence of either recent
binary interactions or star formation. These stars
occasionally expel matter to a centrifugally supported, thin disk, with
a small opening angle, e.g., \citep{CJT15}. The presence
of disks is heralded by emission in the lower Balmer spectral lines.
Our use of the term ``\gc\ stars" is distinct from the definition
given in the General Catalog of Variable Stars, e.g., \citet{S09}, and is
framed within the context of X-ray characteristics, as discussed herein.}
However, with the discovery of anomalously high X-ray flux from the 
direction of this star \citep{Jernigan76, MWS76}, it eventually was revealed
that it is not a typical Be star after all - Be stars at large emit at 
most a few times more soft X-ray flux than normal B stars \citep{Cohen00}, 
ostensibly due to shock interactions in their intermediate latitude hot winds. 
As we will see, the study of the generation of hard X-rays from a class
of so-called ``\gc\ stars" links a wide field of astrophysical subdiscipines. 
In addition, \citet[][```M07'']{Motch07} and \citet[][``N13'']{Nebot13} have 
noted that \gc\ stars contribute very significantly to the X-ray flux emitted 
from Galactic massive stars.  This makes them relevant to an understanding 
of the energy budget of the interstellar medium (ISM) as well as the 
evolutionary consequences
of noncataclysmic expulsions of matter from Be binary systems.

 The saga begins with the discovery that \gc\ is not just an X-ray source 
but a copious emitter of hard X-rays \citep{White82}.
This realization started something of a cottage industry aimed to develop 
an understanding of the mechanisms behind its surprising X-ray properties. 
The initial concept, which had been successfully
applied to the generation of high energy for almost all high mass 
X-ray Be binary stars, is that the X-rays result from synchrotron 
emission (for a strongly magnetic secondary) and/or thermalization of 
accreted matter onto the surface, accretion column, or accretion disk 
hosted by a degenerate companion. In any of these scenarios
the high energy emission ultimately results from the
deep gravitational potential of the degenerate secondary.  White et al. 
professed ``little doubt" that \gc\ is a Be-NS (neutron star) system
similar to the well known X-ray pulsar X\,Per. The reason for this 
judgment was that the light curve showed in their words a ``factor of
2 quasi-periodic variations on a timescale similar to 
the pulsations from X\,Per." Here they anticipated a timescale of
13.9 minutes, which is the pulse period in this X-ray pulsar. 
In addition, White et al. rejected coronal 
and white dwarf (WD) accretion mechanisms. In the latter case they made 
this judgment because the X-ray luminosity was considered too high 
for WD accretion, according to the mass loss rate estimated for \gc.~ 
Despite these arguments it soon became clear that \gc~ is not just another
example of a Be-NS system.

\subsection{Recent discoveries of X-ray properties.}
\label{recdisc}

  By the end of the 20th century new generations of X-ray~ satellites
provided data with improved spectral and time resolution.
These data clarified the X-ray properties of \gc\ and a small number of 
``analog" members of this class and also established that the spectra 
were thermal and optically thin in nature. Also, the amassing of new 
light curves could not corroborate claims of persistent periodicities.
Breakthroughs in evaluating the X-ray characteristics came from the advent 
of the {\it Rossi X-ray Timing Explorer (RXTE)} satellite, with its 
design purpose of providing reliable high time resolution light curves 
through the use of its six detectors comprising the {\it Proportion Counter 
Array (PCA)}, and from the appearance of {\it Chandra} and {\it \xmmn} 
satellites equipped with high resolution spectrometers sensitive to 
soft-to medium band flux (which we define somewhat arbitrarily as
$\ltsim$3\,keV, i.e., $\gtrsim$4\,\AA).

  The new generation light curves demonstrated that flux 
variability occurs on four basic timescales: rapid (seconds to a minute), 
intermediate (a few hours), long (a few months), and episodic (years). 
Meanwhile, results from turn-of-the-century X-ray spectrometers 
have indicated that the X-ray spectra are {\it basically} thermal. 
We clarify the qualifier ``basically" as follows. 
First, one finds the continuum and the Lyman\,$\alpha$ lines of 
 Fe\,XXV (1.85\,\AA) and Fe\,XXVI (1.78\,\AA)
can be fit well to a primary optically thin plasma component described
by a single hot energy temperature, which we will call $kT_\mathrm{hot}$.
Second, the soft X-ray domain X-ray continuum suffers photoelectric 
absorption, which can be modeled by the equivalent of a hydrogen column
density (hereafter $N_\mathrm{H}$), which is mainly attributable to the ISM 
as well as a local one {$N_\mathrm{H_b}$ along the sight line to the hot 
plasma component.
In addition, a number of emission lines in the soft X-ray waveband 
indicate the presence of additional cooler optically thin plasma components. 
If the spectral coverage is limited only to the hard X-ray region the 
continuum alone, one can determine a reasonably accurate value of
$kT_\mathrm{hot}$ -- 
but not necessarily also model the absorption column density. 
This is particularly true if one does not anticipate the possibility of 
multiple absorption columns.
In such cases both the determination of $kT_\mathrm{hot}$ and the description 
of the column absorptions can be badly compromised \citep[][``S04'']{S04}.

  Given these properties, $\gamma$\,Cas has become a compelling 
mystery object with more than enough surprises to attract considerable 
interest within the stellar X-ray community \citep{GN09}.
In the intervening time only some of the answers we now have as to how
and where the hard X-rays are created can be attributed to the latest
generation of X-ray satellites.
Rather,  crucial new information has come with temporal correlations 
of contemporaneous X-ray, optical, and UV data. 

\subsection{General properties.}

\subsubsection{Stellar properties. }

Apart from its X-ray emission, \gc\ appears to be a typical B0.5\,IVe star. 
Its luminosity class IV is consistent with its being a field 
main sequence B star and having an age of $\ge$15---20 Myr. Herein we will
assume that \gc\ has a mass near $15\,M_{\odot}$, a revised Hipparcos distance 
of 168\,pc \citep{vL07}, a radius of $10\,R_{\odot}$, and an 
effective temperature $T_{\mathrm eff}$ = 28\,kK. With these parameters the
uniform stellar disk diameter on the sky corresponds to 0.44 
milliarcseconds \citep{Stee12}.  Robotic observations with the 
16-inch ``T3" Automated Photometric Telescope (APT) using the Johnson $B$ 
and $V$ filters consistently for more than 15 years indicate the presence 
of a robust signature with a period of 1.215811\,${\pm 0.000030}$ days 
\citep[][``HS12'']{HS12}, which is interpreted as the rotational period
of \gc.~ 
Indeed, when one combines the 1.2 day period with the obliquity of the
star/disk system (about 45$^o$) from  Long Baseline Optical Interferometry 
or ``LBOI" \citep{Stee12} one can reconcile the $v\,sin\,i$ value from 
spectral line broadening with the critical velocity at the equator
for a B0.5\,IV star.

\subsubsection{Circumstellar disk measurements.}

  Because its H$\alpha$ line exhibits strong, generally double-peaked,
emission, it is clear that \gc\ hosts a flattened circumstellar 
``decretion" disk. The emission has been present for most all of its nearly
150 year observed history. A chronicle of the variations during the early and 
mid-20th century has been given by \citet{DFS83} and \citet{Harmanec02}. 
These authors noted that the star's H$\alpha$ emission has not always been
detectable, and in fact it disappeared during 1942--1946 after a brief period 
of ``spectacular variations" (SV).  Although many astronomers believe these
variations are due to internal instabilities rising to the Be star's surface,
\citet{Hummel98} has attributed them to the precession of the Be disk
that is tilted with respect to the star's equatorial plane.
However for such an explanation to hold, the orientation of the disk would 
have to change, and thus depart from the rotational plane, 
at the start and end of the SV era. The observational techniques needed 
to determine disk coplanarity were not developed at that time. 

In recent history this emission has been more or less slowly building 
since a sudden outburst in 1969. Documentations of the H$\alpha$ emission 
have been carried out, albeit with heterogeneous instruments since 
1991, and with a single stable low-resolution spectrographic-CCD system 
since late 1994 \citep[][``S12'']{Pollmann14, Smith12}. 
These observations showed the existence of
a quasi-periodicity of the Violet/Red emission ratio extending 
over the last three decades of the 20th century, but this cyclicity 
had disappeared by 2000. 
\citet{SJ07} have published a self-consistent model for the \gc\ disk 
that assumes both radiative-equilibrium and an axisymmetric geometry.
These authors found a volume density at the equatorial disk base
of 3--5$\times$10$^{-11}$ g\,cm$^{-3}$, similar to earlier results by
\citet{MST00}. The latter paper gives a vertical column density at the
base of 0.2 g\,cm$^{-2}$ (10$^{23}$ atoms\,cm$^{-2}$).

  The circumstellar disk of \gc\
has been resolved by a number of LBOI studies. The best models 
for the visibility functions are consistent with the geometry of an
ellipsoid with a Gaussian drop-off in density. 
In addition to LBOI observations centered on the H$\alpha$ line,
observations have been carried out across various continuum bands,
including $V$, ``near H$\alpha$," and infrared 
$H$ and $K$ bands (1.85\,$\mu$m and 2.1\,$\mu$m, respectively).
Brief histories of LBOI studies of this star/disk system 
can be found in \citet{TGV86}, \citet{Quirrenbach97}, \cite{Stee96}, 
\citet{Tycner06}, \citet{Gies07}, and \citet{Stee12}. These studies give 
important constraints on the structure of the inner disk of \gc,~
including its opening angle and density drop-off with radius. These
results also demonstrate that the spatial extent of \gc's disk in the sky
increases at long wavelengths. This is in agreement with the expectation 
that infrared continuum radiation of the disk is dominated by bound-free 
emission and thus become more optically thick at these wavelengths.
\citet{Berio99} used time-resolved LBOI in H$\alpha$ to track 
prograde precession of a one-arm density wave embedded in the disk.
In addition, LBOI observations attached to a high dispersion 
spectrograph have shown that the disk obeys a Keplerian rotation law 
around the Be star out to the outer limits of its detectability
\citep{Meilland11, Stee12}.
Note, however, that these data do not constrain the rotation rate of the
inner disk, that is to within about one stellar radius of the surface
of the Be star.

\subsubsection{Description of binary orbit.}
\label{binry}

   The radial velocity curve for the \gc\ binary system has been determined 
from monitoring the wavelengths of selected points on the H$\alpha$ profile. 
Subtle changes in this profile can produce errors in the shape of the 
RV curve, making it challenging to measure the star's radial velocity
accurately. Nonetheless, it is now clear that the system has an orbital 
period of 203.55$\pm{0.20}$ day, a full velocity amplitude 2K = 
7\,km\,s$^{-1}$, an eccentricity of $\leq$0.03, and a secondary mass
near $0.8\,M_\odot$ \citep{Harmanec00, MBK02, Nem12, Smith12}. 
Given these parameters the orbital separation of the two stellar components
is 36.5\,R$_*$ and the Roche lobe of the Be star is 21.4\,R$_*$, i.e., 
well beyond the observable outer edge of the circumstellar disk. It is
also thought that the outer disk edge is truncated by 3:1 period 
resonances with the  secondary star \citep{Okazaki01}.
The practical effect of this is not to create a sudden cut-off at this
radius but rather to greatly steepen the density distribution beyond this
point \citep{Okazaki02}. From this consideration the accretion of matter 
will be at least somewhat diminished, thereby arguing against an explanation
involving X-ray emission from an accreting degenerate companion..

\section{High Resolution X-ray Spectroscopy}

 We begin our detailed description of the phenomenological properties 
of the X-ray emitting plasmas, 
by discussing the analysis of high resolution spectroscopic data.

\subsection{ Discovery of multiple thermal plasma components.}
\label{dscvr}

  Prior to the launches of {\it Tenma,} {\it Ginga,} and {\it BeppoSax} 
during the period 1983--1996, various studies had lacked the energy 
coverage and resolution to differentiate a thermal
from a power law model in the \gc~ energy spectrum. These instruments as 
well as {\it Chandra} and {\it \xmmn} that followed them in 1999 showed
the dominant plasma component has an energy temperature $kT_\mathrm{hot}$ 
$\approx$ 12--14\,keV  (i.e., 1.4--1.6$\times$10$^{8}$\,K).
Significantly, \citet[][``M86'']{M86} discovered the 
presence of the Fe\,XXV and Fe\,XXVI lines in the X-ray spectrum, which 
although they are blended is important because they strongly suggest the 
spectrum is thermal and optically thin, 
and therefore that it is not formed by accretion onto a NS.

  The first high resolution spectrum of \gc\ was obtained in August 2001, 
through the {\it Chandra} High Energy Transmission Grating spectrometer (S04). 
This spectrum covered the wavelength range 1.6--25\,\AA\ and immediately 
revealed surprises. Foremost among them was its steep gradient with wavelength. 
This was due only partly to the already high expected energy temperature 
of the dominant plasma. The remaining cause of this steep slope was 
the intervening high local column density that was present at this time. 
The low photon count rate near the short wavelength edge and the absorption
column rendered an accurate determination of the temperature uncertain,
so a value of 12.3\,keV was adopted from an earlier analysis of the 
{\it BeppoSax} spectrum \citep{Owens99}. However additional optically thin 
thermal components were needed in the modeling to explain the presence of 
Lyman\,$\alpha$ emission lines arising from even-Z intermediate elements.
In particular, towards longer wavelengths it became necessary to add plasma
components of lower temperatures. 
In all S04 found that four components were required  for a spectrum
obtained in 2001, as S12 also found for spectrum obtained in 2010. 
\citet[``L10"]{L10}
had a slight preference for three components for a spectrum taken in 2004.
The hot component, with $kT_\mathrm{hot}$$\approx$12.3\,keV, 
was found to be responsible for 80--88\% of the X-ray flux in the 
0.2--10\,keV energy band.\footnote{The ``warm" and ``cool" components
have temperatures of typically 2.4--6\,keV and $\sim$1\,keV, respectively.
Throughout this review we use these terms for these energy ranges,
as well as $kT$ $>$7\,keV for ``hot." }
Recent {\it \xmm/EPIC} (European Photon Imaging Camera) spectra,
which provide moderate resolution coverage 
for wavelengths 1--41\,\AA, have pointed to similar $kT_\mathrm{hot}$ 
values, in particular 13.5--15.7\,keV for four observations distributed 
over 40 days in 2010 (S12). However, it is not
clear that the plasma temperature is changed over this short 
timescale because \citet{L10} found a similar range of values,
 $kT_\mathrm{hot}$ = 12.4--14.3\,keV, for a single {\it \xmm} observation made in 
2004, though modeled with different analysis software. In all, while there 
may be some variation of the temperature of the primary plasma component, 
its documented variation is no larger than 2\,keV.

 Additional {\it \xmm} spectra of \gc\ were obtained in February 2004 
(L10) and July--August, 2010 (S12). Importantly, these observations were
not just made through the European Photon Imaging Camera (EPIC) $pn$ 
system (and in the latter case also the MOS1/MOS2 camera).
The two Reflection Grating Spectrometers 
(RGS1/RGS2), which cover emission over 6.2--38\,\AA\ at high resolution. 
Also, the long-wavelength limit
of the RGS system allows generous coverage of the soft X-ray regime. This
means a better modeling of photoelectric absorption columns than 
{\it Chandra/HTEG} as well as the 
inclusion of emission lines arising from hydrogenic CNO ions as well as
 so-called {\it ``rif"} triplets discussed in $\S$\ref{rifdns}. These studies 
also showed a preference for three, or more probably four, optically thin 
plasma components. S12 also found that a continuous Differential Emission
Measure (DEM) model gave a poorer 
fit to the data, 
meaning that the spectrum could indeed be best fit with plasmas having 
3--4 discrete plasmas having different temperatures -- not a single 
broad distribution of temperatures. 
Similar to earlier colorimetric monitoring studies using the 
{\it Rossi X-ray Time Explorer (RXTE)} 
satellite by \citet[][``SRC98'']{SRC98} and \citet[][``RS00'']{RS00}, 
the temperature of the dominant (hot) plasma from $\gamma$\,Cas exhibits 
little epoch-to-epoch variation.
Of the components found for a four component model, the third (having 
$kT_\mathrm{warm,2}$ = 2.5--4\,keV) usually has the smallest 
emission measure. Even so, its presence is essential for the 
formation of Fe\,L shell and other lines. S12 found that the 
temperatures and emission measures of the second and third
(``warm'') components may well be variable on timescales of weeks.

  There is contradictory evidence of whether thermal components alone
are altogether sufficient to describe the X-ray spectrum of ``\gc\ stars.''
In their recent study of \gc, \citet{Shrader15}, combined {\it Suzaku} 
and {\it INTEGRAL} data. This  extended the energy range 
under study to 100\,keV. These authors discovered that a thermal component 
having $kT_\mathrm{hot}$ $\approx$14\,keV is adequate in their modeling 
to fit the data.  The Shrader et al. fit to the continuum is shown in 
Figure\,1 and supports this conclusion. 
However, in their studies of high resolution spectra of \gc,~ 
S04, L10, and S12 found the presence of relatively strong Ly\,$\beta$, 
Ne\,X and O\,VIII lines in the spectrum of \gc\ to be incompatible with a 
thermal model that fit the Fe\,XXV/Fe\,XXVI lines and surrounding continuum.
At the high energy end of the spectrum the hinted existence of a relatively 
strong Fe\,XXVI Ly\,$\beta$ line in the spectrum of the first known and 
brightest \gc\ analog HD\,110432 was first noted on by \citet{TO01}
and discussed further by \citet[][``L07b'']{L07b}.
This feature is unresolved and merits modeling with improved 
description of the line formation mechanisms.

%FIGURE1
\begin{figure}[ht!]
\label{shdr}
\begin{center}
\includegraphics*[width=8.0cm,angle=-90]{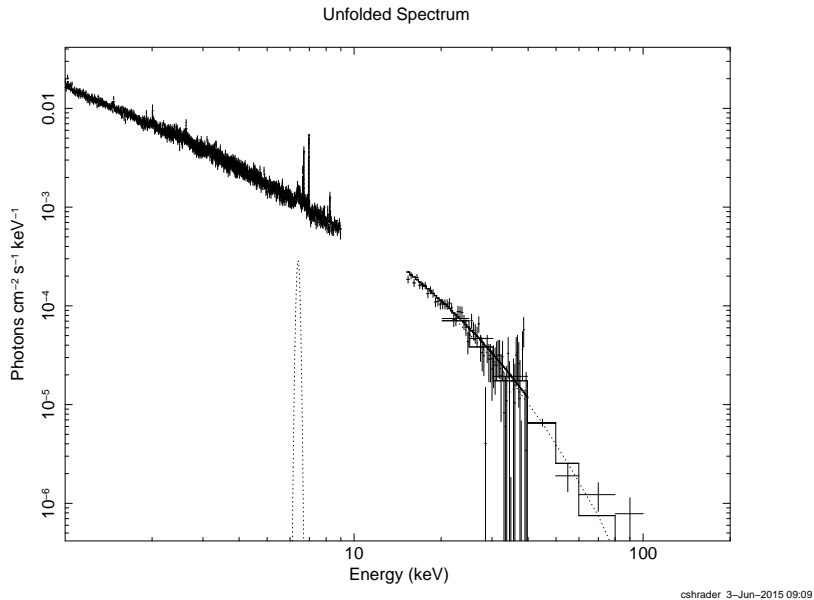}
\end{center}
\vspace*{-0.15in}
\caption{The high energy spectrum of \gc\ according to combined {\it Suzaku} 
and {\it INTEGRAL} observations. The fit is to an optically thin thermal 
model with $k$T = 14.4\,keV and a Gaussian to simulate the Fe\,K fluorescence
feature. With permission of \citet{Shrader15}.}
\end{figure}

  We remark also on the origin of the ``cool" plasma component, which
has an energy temperature $k$T$_{cool}$ $\approx$0.1\,keV. This value 
is roughly consistent with wind-shock plasma temperatures of other 
early B-type stars. However, the volume emission measure is 
a few times larger than shocks caused by B star winds (S04, L07b). 
This fact casts some doubt that we are observing the canonical B-star wind
in this spectrum. Subsequent analyses by L10 and S12 of more recent {\it \xmm} 
soft X-ray spectra has reinforced this doubt, for example because of
their non-negative radial velocities and the short-term variations of
only the soft X-ray flux.
There is some ambiguity here because of the fact that the wind
density varies as a function of latitude among Be stars. Also, the wind 
density for \gc\ has been known to vary with precessional phase of 
the disk-embedded density wave by about an order of magnitude \citep{TK94}.

In our view the question of whether the fourth plasma component is due to
the star's wind is still open because we do not
know whether the X-ray generation mechanism might influence the
generation of winds.

 To unpack the spectroscopic diagnostics revealed by each plasma
component, we exhibit an ``unfolded spectrum" of \gc\ as in Figure\,2. 
The component spectra are computed with the XSPEC modeling program
for an {\it \xmm/RGS} observation on August 2, 2010 (S12). 
The observed spectrum is depicted by a dashed black line at the top. 
Note that the $kT_\mathrm{hot}$ component already almost matches the observed 
continuum spectrum, though not the line spectrum in the long wavelength range 
indicated.
The figure clarifies how the emission line spectrum at longer 
wavelengths imposes the requirement that cooler plasmas be included in
the overall spectrum model.

%FIGURE2
\begin{figure}
\label{unfld}
\begin{center}
\includegraphics*[width=10cm,angle=-90]{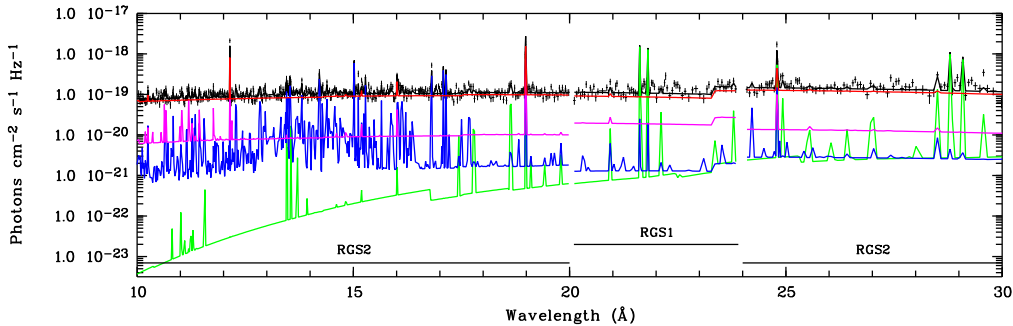}
\end{center}
\vspace*{-2.25in}
\caption{Line- or color-coded ``unfolded" models for the {\it \xmm/RGS} 
spectrum of $\gamma$\,Cas. Horizontal bars give the wavelength coverage 
of the separate RGS1 and RGS2 cameras. 
Sorted by temperature the components are shown as thin continuous 
($kT_\mathrm{hot}$ = 12--14\,keV), thick continuous black 
($kT_\mathrm{warm,2}$ $\sim$2.4\,keV), light dashed 
($kT_\mathrm{warm,1}$ $\sim$0.64\,keV), and thick 
dashed lines ($kT_\mathrm{cool}$  $\sim$0.1\,keV). 
In the on-line version the respective components 
are represented in red, magenta, blue, and green. After \citet{Smith12}.}
\end{figure}

From these models one point should be stressed: {\it even at long 
wavelengths the flux of the hottest component dominates the continuum.}
Then, for example, the attenuation of the soft X-ray flux is the
sign of photoelectric absorption due to an intervening cold gas column
along the line of sight to the hot site(s) situated in background.

\subsection{Two absorption columns toward \gc.}
\label{2col}

  One of the fortuitous aspects of the modeling of these \gc\ data is the
recognition that the broad range of continuum wavelengths 
%covered at high precision 
allows one to determine that two separate absorption
columns to the X-ray sources are required. The density for the
first column, $N_\mathrm{H_a}$, alone is consistent with the UV-derived 
ISM column. Thus, it is assumed to be nonlocal and unrelated to
processes prevailing near the Be star.
The additional column, $N_\mathrm{H_b}$, affects only the
hot plasma component, at least so far as current quality X-ray
spectra allow us to know. It exceeds the ISM value
typically by a large, time-variable factor (2--740$\times$), but it 
covers only one quarter of the hot plasma site(s), a fraction
that S04 expected based on the assumption that the emission sites
are distributed randomly across and close to the Be star's surface and
also on the inclination of \gc\ to our 
line of sight; $\frac{1}{4}$ is the ratio of the visible 
areas in the star's ``Northern" and ``Southern" hemispheres.
The upper part of Figure\,3 depicts for \gc\ the effects of attenuation 
on the soft X-ray region due to the ISM ($N_\mathrm{H_a}$) and 
$N_\mathrm{H_b}$ columns (i.e., as observed), and also flux corrected as if 
only the $N_\mathrm{H_b}$ column were present. 
Additionally, it shows the correction to the flux as if neither column were 
present. The lower part of the figure highlights the effect for the \gc\ 
analog HD\,110432 with the single absorption column, $N_\mathrm{H_a}$, present 
(as observed) and also corrected for the effects of this absorption.
Thus, the highest two flux spectra in this figure (spectra {\it b} and {\it e})
represent what the spectra would look like without any intervening absorption.

%FIGURE3
\begin{figure}[ht!]
\label{gchd}
\begin{center}
\includegraphics*[width=8.0cm,angle=-90]{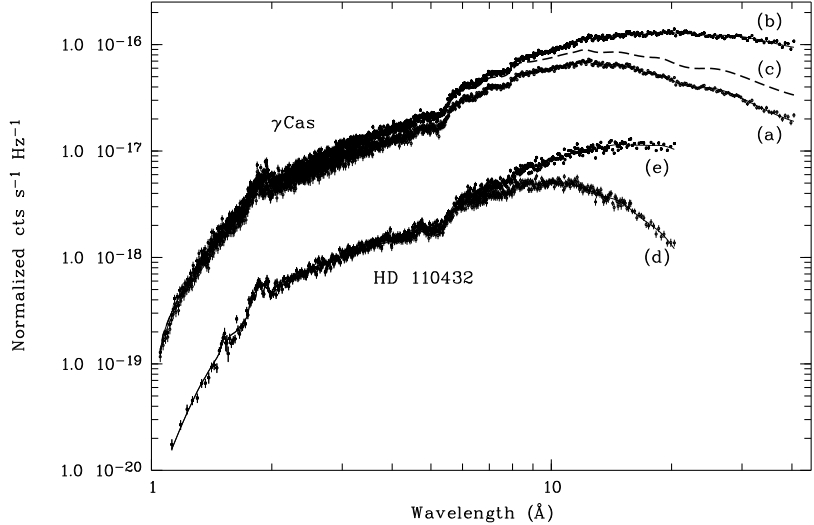}
\end{center}
\vspace*{-0.15in}
\caption{{\it \xmm} spectra of \gc\ (spectra (a-c) and HD\,110432 spectra 
(d-e). Spectra (a) and (d) are the observed spectra on 
August 24, 2010 and July 3, 2004 
for the respective sources. The remaining three are pseudo-spectra representing
the observed spectra corrected for the modeled local ($N_\mathrm{H_b}$) 
and/or ISM ($N_\mathrm{H_a}$) absorption column. Spectrum (b) shows the effect 
of removing both local and ISM absorptions toward \gc,~ whereas spectrum
(c) depicts the spectrum as if only the ISM absorption were removed.
Spectrum (e) represents the removal of the single-column absorption 
toward HD\,110432 due to the ISM. 
}
\end{figure}

\subsection{Spectroscopic measures of the volumetric density.}
\label{rifdns}

   The  density-sensitive {\it rif} (resonance/intercombinational/forbidden)
triplet lines are widely used in X-ray spectroscopy to estimate the 
volumetric density of a hot emitting plasma. For the $\gamma$\,Cas spectrum
the absence of the forbidden component and the near equality of the 
{\it r} and {\it i} line components arising from He-like ions Si\,XIII and
Ne\,IX could itself indicate a high density ($\gtrsim$10$^{11}$\,cm$^{-3}$) 
for the ``warm" plasma components.
However, the implied dominance of collisions setting these ratios could
just as easily be caused by quenching due to collisional deexcitation
due to far-UV radiation from the nearby Be star. 
Therefore, the triplets turn out not to be decisive indicators of the 
local plasma densities. Nonetheless, S12 pointed out
that the observed ratios of certain L-shell Fe ion lines is consistent 
with even higher densities (10$^{13}$--10$^{14}$\,cm$^{-3}$) for the
``warm" components, according to the models of \citet{ML01}. 
The typical density of the 
$kT_\mathrm{hot}$ plasma is at least this high (see $\S$\ref{xrfl}).

\subsection{Correlations with the $N_\mathrm{H_b}$ column.}

  A surprise from these analyses was that the strengths of several 
important spectral lines and their possible correlations with the 
$N_\mathrm{H_b}$ column are evident over time. Of particular note, varying
line strengths are those for which Fe, Ne, and N abundances are derived,
and these often lead to nonsolar abundances. In particular,  
the lines are those from the K-shell ions only of Fe\,XXV 
and Fe\,XXVI and the Ne and N abundances from 
Ly\,$\alpha$ lines of Ne\,X and Ne\,IX and N\,VII and N\,VI ions. 
In contrast to the Fe abundance derived from the L-shell ion lines, 
which is solar-like, the equivalent widths of Fe\,XXV and Fe\,XXVI lines 
are abnormally weak, indicating a low abundance - see Table\,1. 
This particular anomaly was first noticed in the first moderate
resolution {\it Tenma} spectrum (M86), and to varying degrees it has 
betrayed itself in every short wavelength spectrum of \gc\ published since. 

%TABLE1
\begin{table} \caption{Density column (units of 10$^{22}$\,cm$^{-2}$), 
EW (m\AA) of Fe fluorescence, Abundances.}
\label{tbl:prop}        
\footnotesize
\centering           
\begin{tabular}{ccc|crr}            
%\begin{tabular}{cccc|crr}            

\hline                    
\hline\\[-2.2ex]
 Date      &  $n_\mathrm{H_b}$  & EW(FeK$\alpha$) &
Fe & Ne    & N \\
%  &  & (m\AA) &  &  \\
\hline                    

 2001 &  10      &  -19         &  0.10  &      $\sim$1  &  $\sim$1 \\
 2004  &  0.023  &  -10         &  0.12  &     2.63 &  3.96 \\
 2010 & 36-74   &  -35 to -50  &  0.18  &     1.80 &  2.33 \\
% 2001 & 0.75 &  10      &  -19         &  0.10  &      $\sim$1  &  $\sim$1 \\
% 2004 & 0.27 &  0.023  &  -10         &  0.12  &     2.63 &  3.96 \\
% 2010 & 0.74--0.96 &  36-74   &  -35 to -50  &  0.18  &     1.80 &  2.33 \\
\hline                                  
\end{tabular} 
\begin{list}{}{}
%\item ``Fe" refers to Fe abundance from K-shell ion lines.
\item
{\it Notes:}
(1) Errors for [Fe$_K$], [N], and [Ne]: ${\pm 0.02}$, ${\pm 0.75}$, 
and ${\pm 0.28}$, resp.; solar units. 
(2) 2010 abundances are 4-epoch averages.
(3) All spectra analyzed are from the {\it \xmm}~ except the {\it Chandra} 
spectrum in 2001.  
(4) After \citet{Smith12}.

\end{list}
\end{table}
%\vspace*{-0.30in}
%%%%

% \begin{list}{}{}
%\item
%{\it Notes:}
%(1) \citet{White82}, (2) \citet{SB06}, (3) \citet{L06},

\subsection{The FeK fluorescence feature.}
% \label{fekflr}

  In addition to the emission lines noted, the \gc\ spectrum shows the
well known ``FeK$\alpha$" fluorescence feature centered near 1.93\,\AA.~ 
This blended feature is produced by irradiation of hard ($\gtrsim$8\,keV) 
continuum X-rays and the resulting fluorescence $\alpha$ transtion from 
any of several low to intermediate ions of Fe in {\it ``cold"} circumstellar 
gas or by 
%It may be produced either as an outcome of 
scattering of X-rays from the surface of a nearby hot star. 
The feature is typically modeled as an {\it ad hoc} Gaussian from 
which its equivalent width (EW) is determined. 
The precise centroid wavelength (1.931${\pm 0.006}$\,\AA) S12 measured 
for the fluorescent feature is consistent with radiative interactions with 
Fe\,I-FeXVIII ions. This is also the basis for our referring to the fluorescing 
agent as ``cold" gas.  

  \citet[][``G15'']{G15} have 
provided an important clue to the nature of the environment of X-ray 
emitting plasma in \gc\ stars. These authors compared the strengths of
FeK$\alpha$ and recombination lines of Lyman\,$\alpha$ Fe\,XXV and Fe\,XXVI 
for High Mass X-ray B (HMXB) stars at large and \gc\ analogs in particular. 
The latter differ from the former in two important respects. First, only the
well investigated 
\gc\ analogs exhibit visible fluorescence and recombination Fe lines  
-- see also \citet[][``L07a'']{L07a}.  Second, the \gc\ stars do not share 
the HMXB property of exhibiting an inverse correlation between EW(FeK$\alpha$),
that is of the Fe\,XXV and Fe\,XXVI lines and X-ray luminosity.
Thus, the behavior of FeK$\alpha$ sets the two groups apart: the spectra 
of the best known \gc\ stars are unique among high mass X-ray stars by 
exhibiting visible FeK$\alpha$ {\it and} Fe\,XXV and Fe\,XXVI recombination 
lines. This behavior suggests that new analogs can be potentially discovered 
among known Be stars by spectra over a narrow energy range, at least if 
the instrumental detection ability at these energies is sufficient.
In case of a photon dearth the hardness of the continuum combined
with the presence of an unresolved FeK line tells us that dedicated 
follow-up X-ray observations should be undertaken to clarify the results 
of a faint-source survey.

It will be important to confirm these trends with continued observations.
Here we note that the apparent correlation of the 
strength of the FeK fluorescence feature with $N_\mathrm{H_b}$ 
(Table\,1) strengthens the case that the feature is created from emission 
of hard X-ray continuum photons into the line of sight from nearby cold matter.
The abundance results present a puzzle, even though a possible way out is 
that it represents an IFIP (Inverse First Ionization Potential) effect.
IFIP/FIP effects are probably somehow caused by a selective diffusion of 
ions across a magnetic plasma \citep{GN09, Laming04, Laming09}.
The efficacy of this diffusion depends on 
the potentials of their first ionization stages.

\section{X-ray flux and related optical/UV variations}
\label{flxvar}

  In $\S$\ref{recdisc} we stated that the X-ray flux of \gc\ varies
on four basic timescales. We discuss them in order of length of these
timescales.  Because many of these
variations have been well explored for HD\,110432, the 
brightest \gc\ analog, we place the discussion of the light curves of 
this star in this section as well.

\subsection{X-ray ``flares."}
\label{xrfl}

  Starting with studies of the \gc\ X-ray light curve, \citet{Parmar93} 
-- {\it Exosat}) and \citet{Horaguchi94} - {\it Ginga}), it became
clear that ``shots" or flare-like events are present in all light 
curves.\footnote{We loosely defined a flare as a sudden, short-lived 
increase in X-ray flux.  The term does not necessarily connote a localized 
explosive magnetic instability, as on the Sun.} They were found to be 
statistically robust and therefore astrophysical in origin and also to 
exhibit $1/f$-like power spectra.\\

\subsubsection{Light curves and power spectra.}

  The first discovery of a single strong flares in the \gc\ light curve 
was discussed by M86 from a {\it Tenma} light curve. This instrument
was sensitive to the energy range 1.3-10\,keV. 
Observations by the {\it RXTE}, early in its mission when its {\it PCA} 
detectors were optimally efficient, succeeded in pushing the short duration 
(FWHM) limit to 4 seconds (SRC98), where the limit still is today.  
An {\it XMM/EPIC} light curve, covering the energy range 0.3--10\,keV and 
shown in Figure\,4, is instructive in the charcteristics it shows. 
First, we can see the presence of a background or
``basal" envelope, which actually constitutes between 60\% and 
70\% of the total flux  measured by {\it RXTE} within the 
2--12\,keV range. This slowly varying envelope is punctuated by distinct
``flares," whose amplitudes range from barely detectable to twice again
the local basal flux. In general these weak flares are far more numerous
that the strong ones and comprise most of the  X-ray flux attributable to
flares (RSH98, RS00, \citet[][``SLM12'']{SLM12}). Individual flares overlap 
typically $\frac{1}{4}$ of the time, thereby giving the impression in low 
quality time series that they can last as long as a few minutes - these are 
(at best) semi-resolved ``flare aggregates." 

%FIGURE4
\begin{figure}[ht!] 
\label{flarelc}
\begin{center}
\includegraphics*[width=9.5cm,angle=-90]{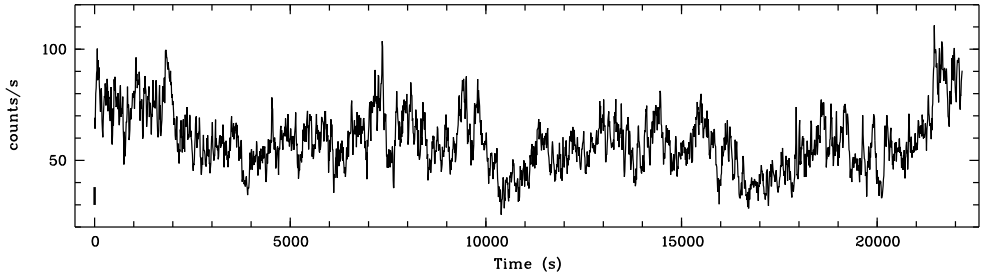}
\end{center}
\vspace*{-2.25in}
% \caption{A stretch of the {\it RXTE} light curve of \gc\ in March 1996. 
\caption{A stretch of the {\it XMM/EPIC} light curve of \gc\ on 2004 
February 5.  The indicated flux error is a typical value determined 
by the data reduction pipeline, in this case for 10\,s binning.
The data show both flares and an underlying basal flux envelope.
}
\end{figure}

 Since these two flux components are roughly equal, they both contribute
important details to a power spectrum of X-ray emission. Figure\,5 represents
the power spectra of two observations of \gc\ two years apart. In these
observations the astrophysical signal can be seen rising from white noise
at high frequencies to greater values at lower frequencies. The spectrum 
is almost linear in a log-log plot.
A sharp eye may notice that the typical slope 
is not quite -1 in the frequency range 10$^{-3}$--10$^{-1}$\,Hz, as one 
expects from ``flicker" $1/f$ noise. It is more correctly described as
``pink noise," which refers to a nonequilibrium-driven dynamical system. 
The plot shows that variations on timescales of several hours dominate
and these will be taken up in the next section. RS00 discovered that details 
in the power spectrum, including the mean slope in the range referred to
above, are sensitive both to changes in the distribution of flare 
strengths and the relative importance of slow changes in the basal flux 
(see $\S$\,\ref{meandr}). 
These factors also contribute to the prominence of a slight break in 
the power spectrum slope at 0.003--0.005 Hz. In particular, a break 
can occur if there are {\it relatively} more strong and long-lived flares 
(or chance unresolved aggregates) than at other epochs.

% FIGURE5
\begin{figure}% [hb!]
\label{ft96}
\begin{center}
\includegraphics*[width=6.5cm,angle=90]{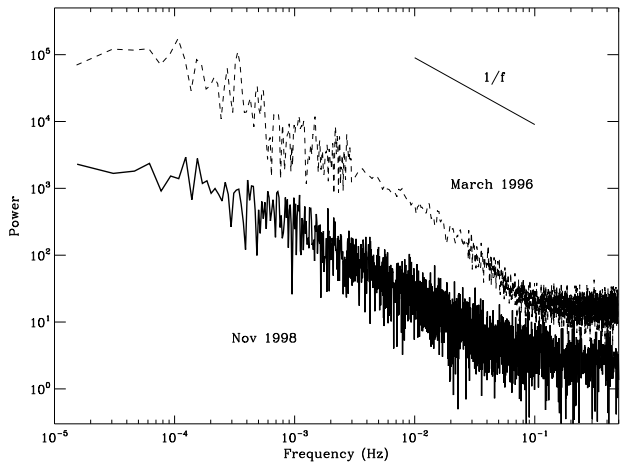}
\end{center}
\vspace*{-0.30in}
\caption{Power spectrum of March, 1996 and November, 1998 {\it RXTE} 
observations (binned to 4\,s). Photon (white) noise is shown in the lower right.
The logarithmic slopes between frequences 10$^{-3}$\,Hz an 
3$\times$10$^{-2}$\,Hz are -1.36${\pm 0.10}$\,Hz and -1.23${\pm 0.13}$\,Hz, 
respectively, i.e., nearly identical but marginally statistically different 
from -1. After \citet{RS00}.}
\end{figure}

\subsubsection{Attributes of individual flares (\gc\ and HD\,110432).}

Several attributes of the resolved flares have been determined for \gc\ and 
HD\,110432. First, in a statistical analysis of a number of short-lived 
flares with high-quality {\it RXTE} light curves, SRC98 found that their 
profiles are symmetrical, suggesting that their decay times are shorter 
than the detectable FWHM lifetime limit of 4\,s.  

Second, the number distributions of flare energies over two different 
epochs have been found to follow exponential functions (RS00, SLM12), 
rather than a power law, as for solar magnetic flares (RS00; SLM12). 
This is true for flares observed in HD\,110432.
On the other hand, SLM12 point out that a power law 
distribution could {\it appear} log-linear if the range of flare energies 
sampled is truncated by instrumental constraints, as is the case for
our sources. 

Third, the upper limit to resolved flare lifetimes is about 150\,s
for \gc\ and about 100\,s for HD\,110432. Most flare lifetimes lie well
within low and high limits, namely 10--30\,s. Mean flare lifetimes and 
energies can be different at different times (SLM12).

Fourth, a comparison of the flare rates of \gc\ and HD\,110432 - using 
both {\it RXTE/PCA} and {\it \xmm} data  showed that once account is 
taken of HD\,110432 being 11 times fainter in the 0.2--12\,keV X-ray 
band than \gc,~ the flare rates of the two stars are equal to well 
within measurement errors (SLM12). 
This rate is 0.009--0.012 flares\,s$^{-1}$. According to studies of
flare rates at different epochs, the variance within this range is 
caused mainly by actual changes in the flare rate with time.

Fifth, SLM12 studied the interval of time between successive flares and
found the distribution was consistent with their occurring randomly.

Statistically, X-ray colors of flares are typically rather close to 
colors of neighboring basal flux. For \gc\ and HD\,110432 small inequalities 
of colors do sometimes occur and with either sign. Generally, the \gc\ 
flares tend to be systematically softer or harder than the basal flux for 
hours or more at a time (RS00). However, it is occasionally the case,
particularly for HD\,110432, that flare to basal colors can reverse
in even several minutes, according to SLM12. These authors found
that 75\% of the time the flare strengths are comparable in soft and hard
X-ray bands. For 15--20\% of events in their study soft X-ray flares 
were stronger than their hard counterparts, while in the remaining 
5-10\% of the cases the reverse was true. These studies demonstrate 
in the main that the flare-producing 
energy reservoirs are fairly constant in terms of their mean injected
energy, though they can at times vary dramatically.

\subsubsection{Differences between flares in \gc\ stars and the Sun.}

  Since we know a great deal about the release of X-rays from 
magnetic activity on the Sun, it is worthwhile comparing the phenomenology
of solar flares with what we observe in the rapid X-ray events in \gc.~
Actually, a
comparison of true solar flares with flare-like events in the light 
curves of \gc\ and HD\,110432 show more differences than similarities.  
An energetic flare on \gc\ can be up to 10$^{33}$ erg\,s$^{-1}$, which
is orders of magnitude more than an energetic solar flare (although in
rare cases solar flares have been observed releasing this energy). 
Flares on the Sun or active lower main sequence stars may last many 
minutes and exhibit slowly decaying tails. 
Solar flares tend to show a power law distribution. As noted, while
the \gc\ flare number distributions seem to show a different law than 
solar flares, it could be illusory. Even so, studies of \gc\ flaring
distributions over time indicate that their slopes vary, and this is not
a data sampling artifact. Nor is it a characteristic of typical solar 
flare distributions.

Another difference with flares in small active regions on the Sun is that 
flares in \gc\ are typically well spaced in time from one another 
 \citet[][``SLM16'']{SLM16}). 
They do not occur as on the Sun and active cool stars in clusters 
or cascades \citep{Aschwanden11}.
Flare cascades are a result of ``Self Organized Criticality," an unstable
condition which can trigger multiple flares locally. In contrast, flares on
\gc~ and HD\,110432 give every indication of being triggered independently
of one another (SLM12). This seems to be a fundamental difference from
true flares on the Sun and related active cool stars. Also, SRC98 pointed
out that the presence of flares is correlated with the introduction of matter
(co-rotating clouds) over areas over the surface.
\\

\subsubsection{\it Determining physical gas properties from flare attributes.}
\label{flrprop}

 In the first statistical analysis of flare properties of \gc,~ SRC98 found 
a strong correlation between the flare (upper) and basal (background) flux 
envelopes as the X-ray light curvesk varied over timescales of an hour or 
longer.  This result, the observation 
that the basal and flare colors are usually almost indistinguishable 
from one another, all combined with the argument that the creation of basal 
flux {\it results} from the flares imply that the flare energy decays 
adiabatically (SRC98). That is, the flare duration is essentially set by
the flare parcel expansion time and not the cooling timescale.
Applying these inferences, and the fact that the X-ray emission is 
optically thin, SRC98 showed that the exploding flare parcels are 
typically small (radius $\sim$5000\,km). Also, {\it they arise 
in plasma densities of $\gtrsim$10$^{14}$ cm$^{-3}$.} This last conclusion 
arises mainly from the observed lifetimes of the most short-lived flares 
and is therefore venue-independent: the density and volume estimates 
derived for the hot plasma sites pertain to whatever the physical 
mechanism for X-ray generation.  This argument reinforces, though is 
much stronger than, results from spectroscopic diagnostics of density 
discussed in $\S$\ref{rifdns}.
\\

\subsubsection{Implications for the generation of hard basal flux.}
\label{hdbas}

  SRC98 investigated the creation of basal X-ray flux for \gc~
as a follow-up to their examination of plasma cooling properties
of X-ray flares.  They found that after an initial injection of 
high energy, gas parcels with typical sizes of a few thousand kilometers, 
expand explosively on  a timescale of 5\,s. Typically some
4\% of their energy is lost during this adiabatic phase.
Eventually radiative cooling dominates the expansion,
but only after further expansion is halted by overpressure from an external 
mechanism.  In view of the presumed role of magnetic fields in enforcing 
corotation of UV-observed clouds, SRC98 attributed the confinement to the 
anchoring of field line loops emanating from the Be star's surface.

During the post-flare cooling stage one should expect to detect emission 
from a continuum of temperatures if the gas expansion continues unabated. 
L10 and S12 tested the idea of {\it in situ} cooling by searching for the
presence of a DEM, that is by plasma emission 
over a continuous range of temperatures instead of one or more discrete ones. 
Their work showed that the  DEM description did not result in a model that
gave an acceptable fit to the X-ray emission line spectrum and thus argued 
against a continuous and unconstrained  plasma cooling.

In their discussion of the generation of basal flux, SRC98 adopted 
a characteristic volumetric density of 10$^{11}$\,cm$^{-3}$ based on the 
timescale of changes in the basal flux and also the frequency where the power
spectrum tends to departs from a linear $1/f$ relation (Figure\,5). 
Using this value and using flare-canopy plasma models developed for solar
flare regions, these authors found that an individual flare expands into 
a canopy with projected area of $\sim$0.002 of the star's and a height
of $\sim$0.05R$_*$ above the star's surface. 
These numbers are self-consistent in the sense that they predict the 
generation of the observed X-ray emission measure of 
$\sim$3$\times$10$^{55}$ cm$^{-3}$. They also assume that at any 
moment we observe a total of 10--20 flare and basal emitting regions. 
SRC98 estimated the mean magnetic fields within these confined regions to 
be $\sim$200\,gauss.
We stress that these estimates ultimately hinge on the assumption that the
immediate post-flare expansion is abruptly braked within a magnetically 
confined volume. The result is that the basal and flare temperatures are 
about equal.  In $\S$\,\ref{msddesc} we will discuss initial conditions 
that might lead to the creation of surface flares.

\subsection{Where do the warm plasmas reside?}
\label{whrwrm}

  The least understood of the X-ray emitting components in \gc\ and
HD\,110432 are the one or two so-called warm plasmas (defined again with
$kT_\mathrm{hot}$$\approx$2.5--6\,keV). This is because these components
are visible mainly through their line spectra; their continuum
emission can be interpreted only roughly through an X-ray colors defined
across short and long wavelength subdomains. A few studies 
\citep{SRC98, L07b,RS00}  
comparing rapid time series behavior between soft and hard X-ray light
curves indicate that they often do not coincide, and this has been taken
as evidence that the warm plasmas are not cospatial with the hard X-ray sites. 
As noted in $\S$\ref{rifdns}, the {\it rif} ratios arising from He-like 
ions do not advance the issue.  The observation of certain Fe\,L lines noted 
in the same previous discussion could lead to densities as low as 10$^{13}$
cm$^{-3}$. This is consistent with the base density of the disk. 
  
  Another interesting diagnostic is the local turbulent velocity measured 
from the (symmetric) lines arising from the warm plasma.  This broadening is 
of the order of 500\,km\,s$^{-1}$ \citep[][]{S04, Smith12}, although in one
case, possibly an outlier, a value as high as 950\,km\,s$^{-1}$ was reported.
The salient point here is that velocities derived from warm plasma lines 
are inconsistent with the higher values of thermal broadening of the hot
plasma ($\gtrsim$1500\,km\,s$^{-1}$; SRC98). 

  In addition, in their {\it HST/GHRS} time series spectra
SR99 found no correlation between the presence of hot plasma
lines of Fe\,V shifted by the same velocities at which Si\,III 1417\,\AA~ 
(a warm plasma line) features were found.

These discoveries validate the conclusion from the X-ray color curves noted
just above that the warm plasmas are in emitted in another place -- even if
the their formations are associated (that is, generally correlated) with
one another.

  S04 speculated that the warm plasma could be created either by
collisions of hot plasma, or perhaps by an initial accelerated
electron beam, into disk material or as an attribute of the slowing of
the plasma cooling as the losses move through plateauing cooling curves 
due to certain dominant metallic ions in the plasma.  However, the improved
accuracy of the component determinations in the \citet{Smith12} study also
indicates the absence of plasma emission between the emitting components.
This suggests to us by now that the continuously cooling scenario is a 
less likely explanation.  

 As for the companion cool ($kT_\mathrm{cool}$) plasma component, in
addition to the arguments mentioned in $\S$\ref{dscvr} one should
emphasize the fact that this component shows lines leading to solar
(at least for [C]) abundances indicates that the soft spectrum is formed 
in an optically thin plasma.

In general, it must be said that the location of the warm and cool
plasmas and their role in local energetics near the Be star are not yet
understood. The high throughput of the envisaged Athena X-ray satellite,
and in particular the high spectral resolution capabilities of its planned
X-IFU instrument, will allow exquisite time-resolved spectroscopy. Line 
ratios, profiles and velocities will provide key diagnostics on the location 
and properties of the hot, warm and soft thin thermal emitting regions.

In addition, future observations with a larger aperture X-ray telescope
can tell us whether the emission lines emitted by the warm plasma are
attenuated or not by the N$_{H_b}$ column, which so far is defined mainly
for the hot plasma sites in the \gc\ spectrum. If N$_{H_b}$ affects the 
warm sites too they are probably be located near the hot sites, even if 
they are not cospatial with them.

\subsection{Quasi-random undulations of the X-ray and UV light curves.}
\label{meandr}

  Using the then novel capabilities of the {\it RXTE/PCA} and {\it ASCA} 
detectors, SRC98 and \citet[][``K98'']{K98}, respectively, highlighted 
quasi-random undulations in the X-ray light curve of \gc\ on timescales of 
2--3 hours. Examples of these variations are exhibited in Figure\,6 
and are
responsible for much of the X-ray power at frequencies $\ltsim$10$^{-3}$\,Hz.
These particular {\it RXTE} observations were unique in that they were 
executed simultaneously with the Goddard High Resolution Spectrograph 
(GHRS) aboard the {\it Hubble Space Telescope} on a 21\,hour campaign in 
1996 (hereafter the ``1996 campaign"). The GHRS spectra were centered
on the Si\,IV line multiplet at $\lambda$1394 and $\lambda$1403.
These spectra were binned to 60\,s. 
Next the fluxes were coadded over a stretch of quasi-continuum 
wavelengths to form a UV light curve \citep{SRH98}. This light
curve showed a series of striking correlations or anticorrelations 
with the X-ray variations during this duration. One of these relations 
was the anticorrelation of the {\it RXTE} fluxes and the UV light curve. 
The latter showed dips of 1--2\% depth corresponding to X-ray maxima. 
These maxima were accompanied by increases in the basal flux. This
suggested that the increased X-ray flux is associated with intervening
structures that absorb UV light from the Be star's surface.

%FIGURE6
\begin{figure}[ht!]
\label{hrsxte}
\begin{center}
\includegraphics*[width=7.0cm,angle=90]{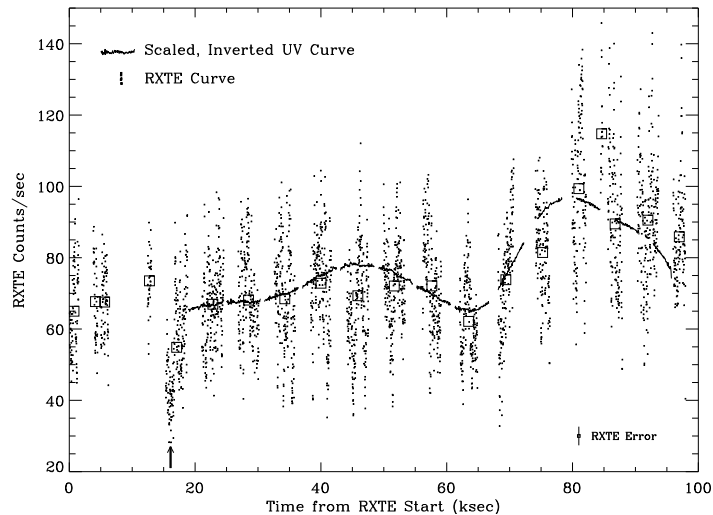}
\end{center}
\vspace*{-0.15in}
\caption{Simultaneous {\it RXTE} light and HST/GHRS light 
curves of \gc\ obtained on March 14-15, 1996. GHRS spectra 
have been binned in wavelength to form a ``UV continuum" light 
curve. The latter curve has been inverted and scaled to match 
variations in the {\it RXTE} light curve. The X-ray variations are 
emphasized by orbital averages (open squares).
The X-ray data are binned to 16\,s; their high frequency ``noise" is due 
to flaring. The upward arrow in the lower left exemplifies 
an X-ray ``cessation" (see text). After \citet{SRC98}.}
\end{figure}

  RSH02 first attempted to model 
the UV continuum dips by supposing that they were caused by rotationally 
advected black spots across the Be star's surface but were unable to 
do so.  The problem was that the variations are too short-lived to 
match the durations of these features. These authors were able to
match the light curves with elevated translucent ``clouds" forced into
corotation above the star. The typical sizes and elevations of these 
structures was found to be 0.2--0.3R$_{*}$, on the assumption of forced 
corotation and that the blobs cross move parallel to the equatorial plane.
The determined elevation of these clouds
allows the transits to be brief enough to match the duration of the 
%light curve dips. Figure\,\ref{uvlcmod} exhibits the best fits to a single 
light curve dips. Figure\,7 exhibits the best fits to a single 
advected surface spot (unsuccessfully) and successfully to three elevated 
clouds.

%FIGURE7
\begin{figure}
\label{uvlcmod}
\begin{center}
\includegraphics*[width=10.0cm,angle=00]{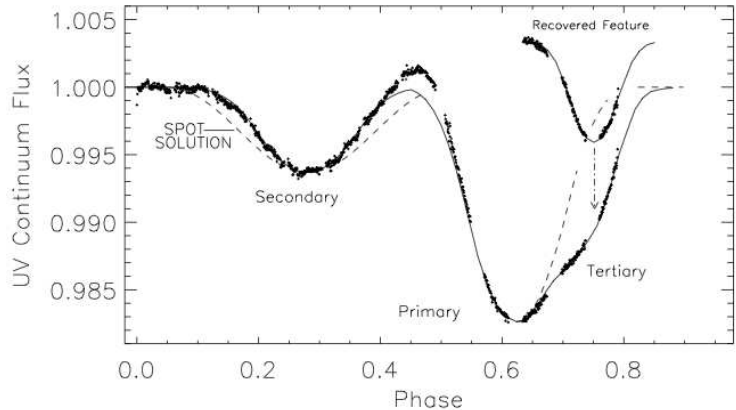}
\end{center}
\vspace*{-0.15in}
\caption{The March 14-15, 1996 UV continuum light curve of \gc.~
This is fit with models of a rotationally advected  surface spot (dashed 
line) and three clouds forced into corotation with radii of $\approx$0.3R$_*$ 
and elevated by a similar distance above the star's surface (solid line). 
The dashed line is the closest fit possible for the spot model and is 
still inadmissible.  After \citet{RS00}.}
\end{figure}

  In addition to the March, 1996 {\it HST/RXTE} campaign, 
\citet[``SR99'']{SR99} also undertook a 33 hour-long (with short
interruptions) high-resolution {\it IUE} campaign on \gc\ in January
1996. They again constructed light curves of mean fluxes in each
echelle order and found features in the time series they identified 
as similar dips to those found in the {\it GHRS} time series.
They discovered that the observed dip
minima grew monotonically stronger with decreasing wavelength, 
that is across the  {\it IUE} echelle orders.
Applying blanketed model atmosphere spectral syntheses, they were able
to show that this progression with wavelength could be explained 
by translucent clouds with a temperature of $\ltsim$10\,kK, which 
is very different from the temperature expected in low density gas  
occupying the outer fringes of the Be star photosphere.
Note that although this temperature is in the expected range of matter in
intermediate or outer regions of the Be disk,
we do not believe the dips originate from matter in the disk. 
The disk matter rotates around the star at a Keplerian rate, or at
$\approx$1R$_*$ nearly so. Therefore any putative stable 
blobs constrained in the disk itself would transit the stellar disk
 {\it too slowly} to be consistent with the dips in Figure\,7. 
We note in passing that given these trends with wavelength from {\it IUE} 
spectra, it is not surprising that corresponding dips have not been discovered 
in the optical light curves of \gc,~ as noted by \citet[][``SHV06'']{SHV06}.

\subsection{Migrating subfeatures line profiles of \gc\ and HD\,110432.}
\label{msfsec}

 \citet[]{Yang88} discovered the presence of ``migrating subfeatures," or
{\it msf,} in spectral lines of the optical spectrum of \gc." These appear 
in the grayscales of timeseries of spectra. These are narrow absorptions 
moving steadily from the blue to the red edge of a rotationally broadened 
line profile, often to be followed several hours later by a new
{\it msf.} The existence of {\it msf} was initially discovered in an
unrelated type of star, a magnetically active young K5\,V star known as
AB\,Dor \citep{Cameron89}.

{\it Msf} cannot arise from rotationally-advected black spots, as on the 
Sun, for several reasons. For example, their motion across the line profiles 
would indicate, as for the UV-absorbing larger ``clouds" discussed above, 
super-critical stellar rotation. The features occur perhaps at 
irregular intervals during some nights of observation and sometimes 
not at all \citep{Smith95}. Because of their 
erratic appearances and their narrow subprofiles compared to the
intervals between them, they are not consistent with the more smoothly
varying (and generally evenly spaced) ``wiggles" in line profiles that
are characteristic of surface nonradial pulsations in other B stars. 
The {\it HST/GHRS} campaign in March, 1996 that we have noted included 
weak lines located near the Si\,IV doublet at 1394\,\AA\ and 1403\,\AA.~ 
SRH98 constructed a grayscale time series of 1280 spectra, each binned over
1 minute. Figure\,8 exhibits a portion of their grayscale in the range 
1404--1417\,\AA\ (where few effects from variations of the nearby 
Si\,IV resonance lines occur). 

%FIGURE8
\begin{figure}[ht!]
\label{uvmsf}
\begin{center}
\includegraphics*[width=10.0cm,angle=90]{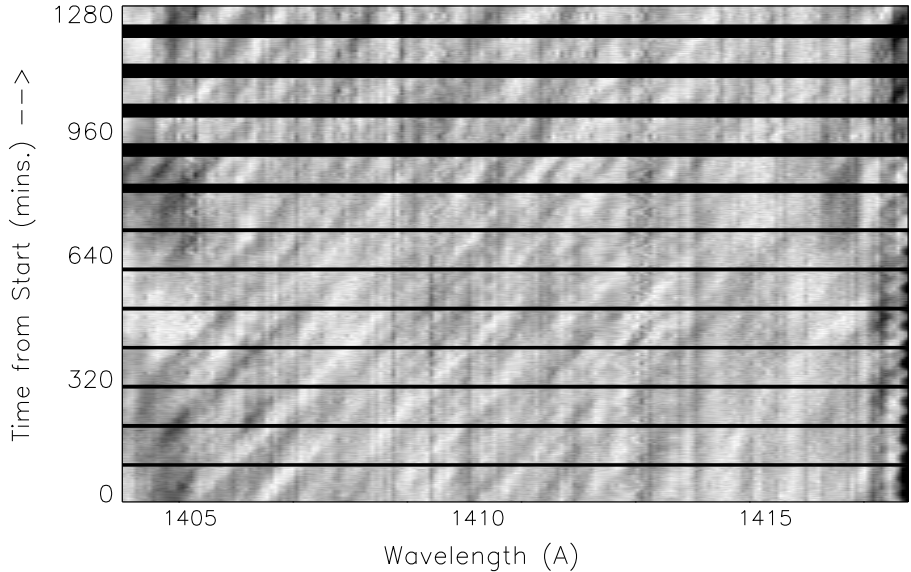}
\end{center}
\vspace*{-0.90in}
\caption{The 1404-1416.5\,\AA\ spectrum of \gc\ during the 1996 HST/GHRS
campaign, which consists of 1280 spectra each binned over a minute.  
The dark diagonal striations moving upward and to the right
are the {\it migrating subfeatures} in the spectrum of many faint spectral
lines. These {\it msf} arise from the Doppler imaging of small translucent 
clouds in the foreground. Individual features last no longer than one to a
few hours (2 orbital segments of the time series). The limits of a
rotationally broadened line profile at these wavelengths is about two 
\AA ngstroms.
The rate of migration of these features across the underlying
line profiles is best explained
by their forced corotation over the star's surface.
Modified from \citet{SRH98}.}
\end{figure}

Dominating this figure is a pattern of diagonal striations
running from the lower left (short wavelengths, starting spectra) to 
upper right. Their mean acceleration is +95\,km\,s$^{-2}$, close to values 
found for {\it msf} in optical spectra of \gc\ \citep{Smith95}. 
These features were found by \citet{Yang88} and \citet{Smith95} in optical
lines of the \gc\ spectrum with very nearly the same acceleration rates.
This fact suggests that, similar to the UV continuum dips discussed above, 
these features arise from cloudlets,
that are forced into corotation a few tenths of a stellar radius 
(at most) above the Be star's surface. The inference that they are elevated 
comes largely from the fact that their acceleration rate is slightly too 
fast to be due to surface advection of spots. 
The {\it msf} striations in our figure can be shown to arise from $\sim$10 
spectral lines identified from Fig.\,4b of SR99 and thus their local
velocities are known. 
%A plot of the 
%mean variations (r.m.s.) with wavelength enables one to define the positions 
%of 10 lines in the UV spectrum (e.g., Figure\,4b of SR99). 

A close inspection of Figure\,8 discloses that the strengths of these 
{\it msf} are quite variable.  In fact, no individual {\it msf} lasts for 
longer than one to a few hours, which suggests 
an environment in which small cloudlets are constantly 
forming and quickly disappearing.\footnote{The rationale for calling the
structures responsible for {\it msf} {\it small,} as opposed to the larger 
clouds responsible for dips in UV light curves, is twofold: 
(1) their existence is more ephemeral (small structures can more easily
dissipate) and (2) a structure of a given size, having properties 
distinguishing it from the background photosphere, is generally much more
visible in a spectrum than the continuum.} 
It may or may not be coincidental that the timescales for the changes
of {\it msf} strengths and basal flux are the same, $\sim$1000\,s.
It should be noted that this type of ephemeral behavior is not
generally descriptive of the circumstellar environment of other Be stars
-- nor does it describe the behavior of matter in the inner parts of 
their decretion disks. 

It is important to add that \citet[][``SB06'']{SB06} published time series 
spectra of the He\,I 6678\,\AA\ line HD\,110432 that also showed 
likely {\it msf on two of six ground-based observing nights.} 
%This is the only study todate of {\it msf} in optical spectra of a \gc\ analog.
To sum up, {\it msf} have been reasonably well documented in 
spectra of only three stars: the object in which they were discovered
(AB\,Dor), \gc,~ and the \gc\ analog HD\,110432.  Serial high-resolution
spectra have not yet been observed in other analogs to see if they
are a universal property of the class.

One may contrast the {\it msf} with the appearance of moving bumps in 
spectral absorption line profiles of unrelated Bp or Be stars.  
These features arise from a vectorial sum of local projected rotational 
and nonradially pulsational (NRP) velocities in a star's photosphere.  
Bumps from nonradial pulsations are common in line profiles of Be stars and, 
except for interferences among comparable amplitude modes, occur regularly. 
(Thus, even if bump amplitudes transit the line profiles irregularly the 
quasi-periodicity of the NRP disturbance is evident in periodograms of light 
curves.) In addition, whenever these NRP features are visible in line profiles 
they represent surface waves generally traveling across the visible hemisphere. 
Line profile bumps are also often present (sometimes as emission as well
as absorption features) in spectra of magnetic Bp stars, corresponding 
to accumulated matter in the two intersections of the magnetic and 
rotational equators.  Such features also appear at regular intervals.
Both of these examples are phenomenologically distinct from the {\it msf}
in \gc, a fact that stems from their
formation in different locations relative to the star's surface.  
 
One unique report by \citet{HV1996} is of moving ``narrow optical absorption 
components" (NOAC) through optical line profiles of spectra or 48\,Lib 
(B3IV\,e-sh).
Although these features exhibit a few similarities with the {\it msf,} 
they also have important dissimilarities. The latter include: (1) the NOAC 
are present only in low excitation metallic lines (generally recognized 
as ``shell" lines), (2) their spectral widths are consistent with only 
thermal broadening, and (3) they occur only within a narrow radial veocity
range of line center.  Like the authors themselves, we do not know the
physical origin of the NOAC features, other than they clearly are formed
in orbiting shell clumps orbiting 10's of radii from the star. 
Nonetheless, the phenomenological {\it dissimilarities} argue that the NOAC
have a different origin from the {\it msf}. It would be edifying to find them 
in the spectra of any other Be shell star, \gc\ analog or otherwise.

\subsection{Further intermediate X-ray/UV correlations in the 1996 campaign.}
\label{xruv}

  In addition to the correlation between X-ray and 
{\it GHRS}-observed UV continuum fluxes observed of \gc\ in 1996 
(Figure\,6), various UV line strengths were found to be 
associated with the X-ray variations. Some of these are documented in 
Figure\,9. {\it This figure shows that when the X-ray flux increases the 
strengths of Fe\,V absorption lines weaken (hence their fluxes increase), 
whereas at the same time the Si\,III and Si\,IV lines strengthen.} 
SR99 found that Si\,III--SiIV and Fe\,V lines are expected to be formed in
matter located just above the surface of a B0.5 star.\footnote{The spectral 
line synthesis models suggest that Fe\,V lines are formed in an ambient 
gas with temperature of 34--45${\pm 3}$\,kK.  This is higher than the 
$T_\mathrm{eff}$ $\sim$28\,kK for \gc.~} 
The parameters of circumstellar gas are such that
irradiation by nearby X-rays is either increased or decreased, respectively.

%FIGURE9
\begin{figure}[ht!]
\label{uvlcmod}
\begin{center}
\includegraphics*[width=10.0cm,angle=00]{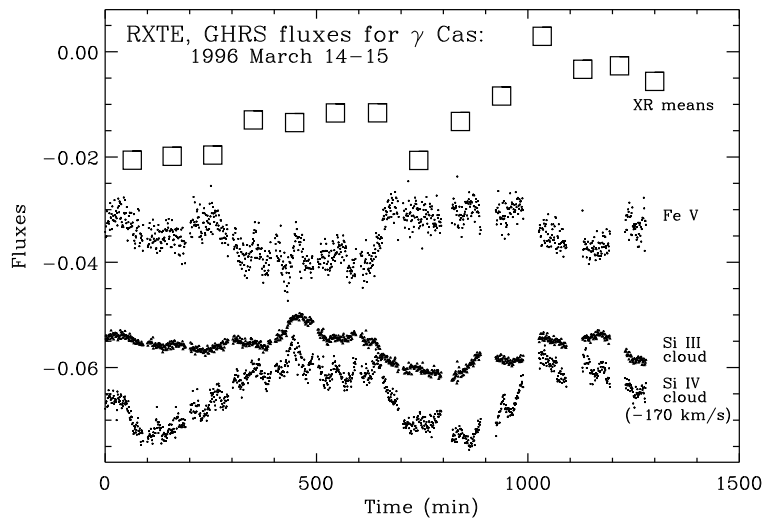}
\end{center}
\vspace*{-0.15in}
\caption{Variations of UV line fluxes against X-ray fluxes during the
GHRS and X-ray campaign in March 1996. The ordinate for the UV lines is
expressed in units of the local continuum flux. The rescaled X-ray flux 
averages (open squares) are taken from Figure\,6 and reflect a 
change of a factor of 2.  After \citet{SR03}.}
\end{figure}

 SR99 discovered correlations of X-ray flux with other UV line 
diagnostics in their 1996 {\it GHRS} observations as well. 
For example, the 1412.9\,\AA\ line, attributed to the underionized stage 
Fe$^{+}$, exhibited a similar anticorrelation, as did the S\,IV 1404.7\,\AA\ 
line and several other features arising from less excited ions.
% and also exhibiting {\it msf.}

Surprisingly, and unlike most neighboring weak lines, the nearby UV 
resonance Si\,IV lines exhibited no internal migration patterns at all, 
even though SR99 were able to simulate this behavior in a nearby S\,IV line. 
The authors explained this unexpected behavior by the addition of a 
separate optically thin column to their spectral line synthesis model.
Thus, their modeling of these UV lines indicated the presence of {\it two}
columns: while most of the lines of sight were optically thick in the 
center of the spectral line,  it is required that a smaller number
of sight lines be optically thin.

\subsection{Claimed periodicities on timescales of hours.}

Early in the exploration of the X-ray properties of \gc,~ astronomers
searched for periods in the expectation that robust X-ray pulsing would 
be observed that would clearly indicate that \gc\ is a magnetized rotating
Be-pulsar or perhaps a highly magnetized Be-WD system. 
However, these hopes have not been realized.
\citet{Frontera87} reported the existence of ``pulses" with a 
period of about 6\,ks based on the observation of four equally 
spaced maxima in the {\it EXOSAT} light curve. 
\citet{Parmar93} reexamined the Frontera et al. data and
concluded that the pulses arose from a fluctuation in one component of 
an extended red noise in their power spectrum.
They then acquired a new longer {\it EXOSAT} observation but were still
unable to detect again a 6\,ks signal. \citet{Horaguchi94} were likewise
unable to find this period or any other period up to 13\,ks from 
additional {\it Ginga} observations. These authors also searched for
correlations between variations in the X-ray and UV and optical bands. 
Their data were based on broadly contemporaneous, not simultaneous, 
time series and were not able to find correlations.

  \citet{Haberl} reported the possible existence of a 8.1\,ks period 
in a soft X-ray light curves from 4 clusters of observations distributed 
over 65\,ks.  They then examined archival {\it EXOSAT} data and could 
not confirm this signal. Due to the limitations of these data, however, 
they did not believe this was a decisive test. The many X-ray light curves 
of \gc\ determined in the meantime have not been able to find this period.
In addition, this report may be criticized because of the poor time
sampling for this short period.

  \citet{TO01} noticed a sinusoidal variation in an 18.7 ks-long
{\it BeppoSax} light curve of HD\,110432. On this basis they claimed 
the existence of a $\sim$14\,ks ``pulse period." However, SB06 demonstrated 
that irregular undulations with a similar timescale also occur in
the X-ray light curve of \gc, and yet it shows no periodicities.
Subsequent light curves of HD110432 have not shown this periodicity but
have shown other such undulations, e.g., \citet{L07b}.

In addition to these claims \citet[][``L06'']{L06} discovered a 3.2\,ks 
modulation in a light curve of the analog HD\,161103, acquired in 2004
with the {\it \xmm},~ by identifying 4 
(out of a possible 5) maxima. However, such a modulation was not present
in subsequent observations carried out in 2012 with the same instrument 
\citep{Martinez14}. 
%were unable to confirm this result.
%These authors find HD161 to have a variable flux & $k$T - 8 yrs apart.

Just as interesting perhaps is the discovery of periodic ``near-cessations" 
in either basal or flare X-ray flux of \gc,~ each lasting
just a few minutes. RS00 first report the existence
of 7--7.5\,hour cessations.  An example 
of a near-cessation is denoted by an upward arrow in Figure\,6.
These authors did not claim that the cessations are strictly periodic. 
In fact subsequent light curves have shown such events to occur during 
some epochs every 7\,hours, but occasionally at other time every 
3.5\,hours and 5.8\,hours \citep[][``RSH02'']{RSH02}. 
Interestingly, from
archival {\it International Ultraviolet Explorer (IUE)} data 
\citet[][``CSR00'']{CSR00} have found a ``flutter" in the fluxes of the 
so called Discrete Absorption Components situated in the far blue wings 
of the C\,IV and Si\,IV resonance lines. These recurred with a 7.5\,hours. 
regularity during an {\it IUE} campaign in January 1982. 

Separately, the occurrence of flux cessations hints at the existence of
an input energy reservoir which empties over a relaxation cycle (SLM12).

\subsection{The 70-day cyclical and annualized signals.}

   Early in the history of monitoring \gc\ in optical and X-ray 
wavebands, it became clear that the light curves vary on timescales 
of more than a day but shorter than a year. 
RSH02 first noticed that optical variations 
occur that are almost periodic, with an average cycle length of 
$\sim$70 days and a full amplitude of 0.02--0.03\,magnitudes in the $V$ 
band. HS12 demonstrated that the cycle properties vary across various 
degrees of parameter space. Thus,
the full range of the cycle lengths is about 50--91 days. 
The cycles can change their lengths and amplitudes from one cycle to
the next. They can even decay and grow anew in less than a few weeks.
In addition, the ratio of amplitudes $\Delta$B/$\Delta$V is variable 
over the years and may have a bimodal distribution.

RSH02 also found, if suitably scaled X-ray variations of \gc\ from six epochs, 
that the optical cycles exhibited a very good relation with epochal averages 
of X-ray flux. This is evident according to Figure\,10), at least during 
this time interval. Using the 
scaling factor required to match the X-ray and optical variations, SHV were 
able to anticipate the scale and increase of the X-ray flux over the course 
of 10 days a few years later. 
Analyzing 15 years of simultaneous {\it ASM/RXTE} X-ray data and $V$ band 
APT observations, MLS15 found that on many occasions the regularly sampled 
X-ray and optical light curves were very well correlated on time scales of 
several months (see their Fig. 4) while only imperfect correlation was 
seen during other observing seasons.
They noted that the correlation could be improved by broadening the range 
of timescales from tens of days to several months and applying the same 
X-ray/optical (disk only) scale factor RSH02 had found to variations over 
the broader timescale of 70\,days to a few years.
The correlation for these wavelength regimes for
annualized variation-averages is shown for 9 years in Figure\,11, left panel. 
The optical averages come from the ground-based APT system while the X-ray
averages come from the {\it All Sky Monitor (ASM)} experiment carried out
by the {\it RXTE.} Figure\,11, right panel, shows the corresponding running 
averages for the two systems, this time for 15 years, showing that well 
correlated variations occur on timescales of a year and longer. 

%FIGURE10
\begin{figure}[ht!]
\label{longper}
\begin{center}
\includegraphics*[width=10.0cm,angle=90]{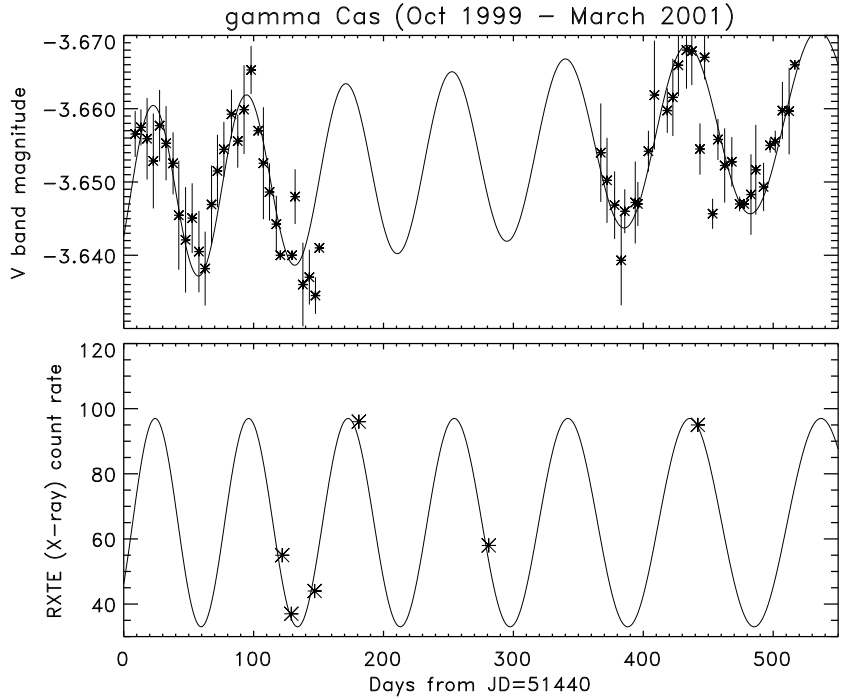}
\end{center}
\vspace*{-0.15in}
\caption{{\it Top:} A comparison of APT $V$ band observations of \gc\
with time during 1999--2001 fit to a modified sine curve (solid line).
{\it Bottom:} Trend-removed sine curve fit from top panel to six
epochal {\it RXTE} X-ray observations. After \citet{RSH02}.}
\end{figure}

%FIGURE11
\begin{figure}[ht!]
\label{asm}
\begin{center}
\includegraphics*[width=4.7cm,angle=-90]{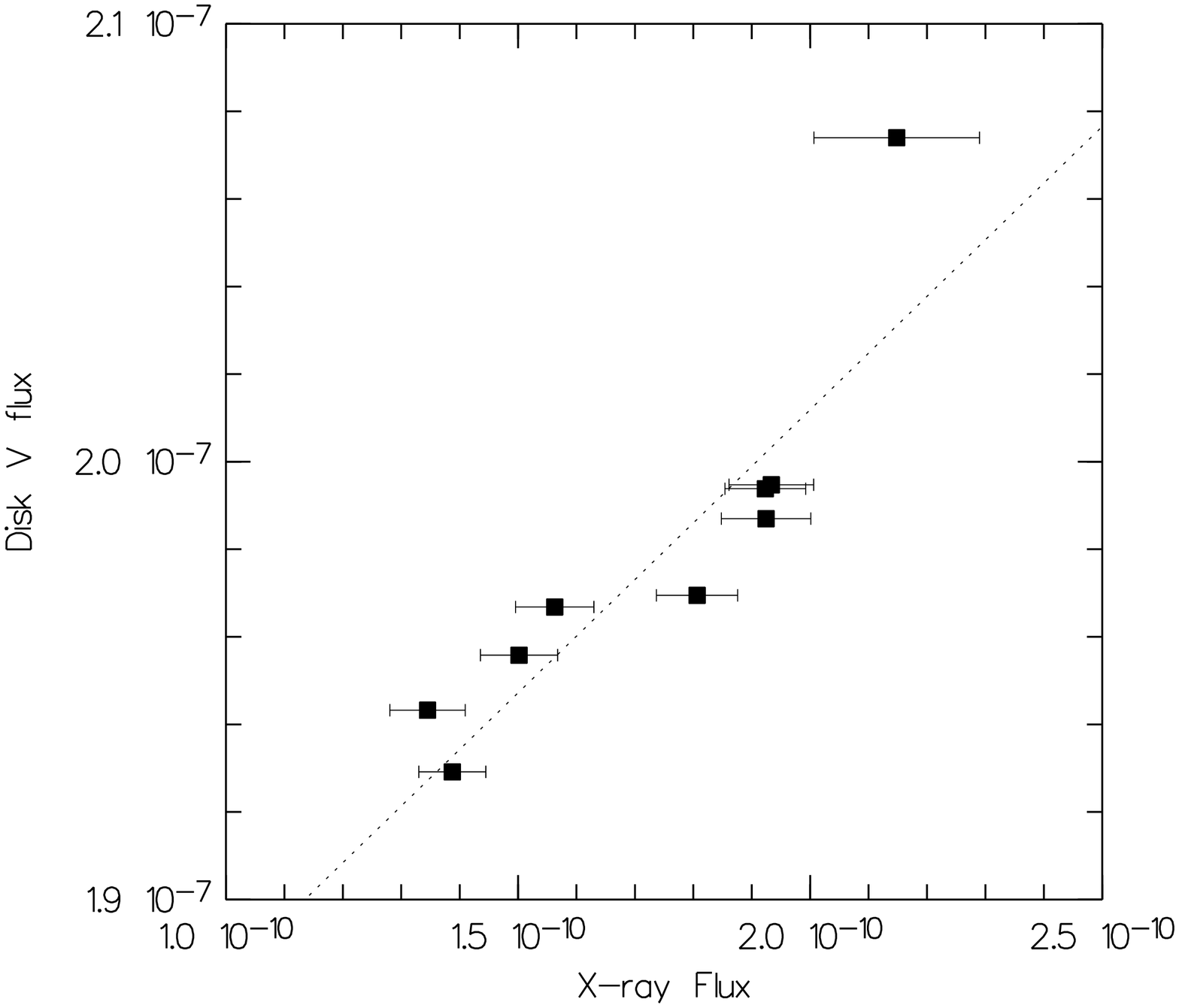}
\includegraphics*[width=4.7cm,angle=-90]{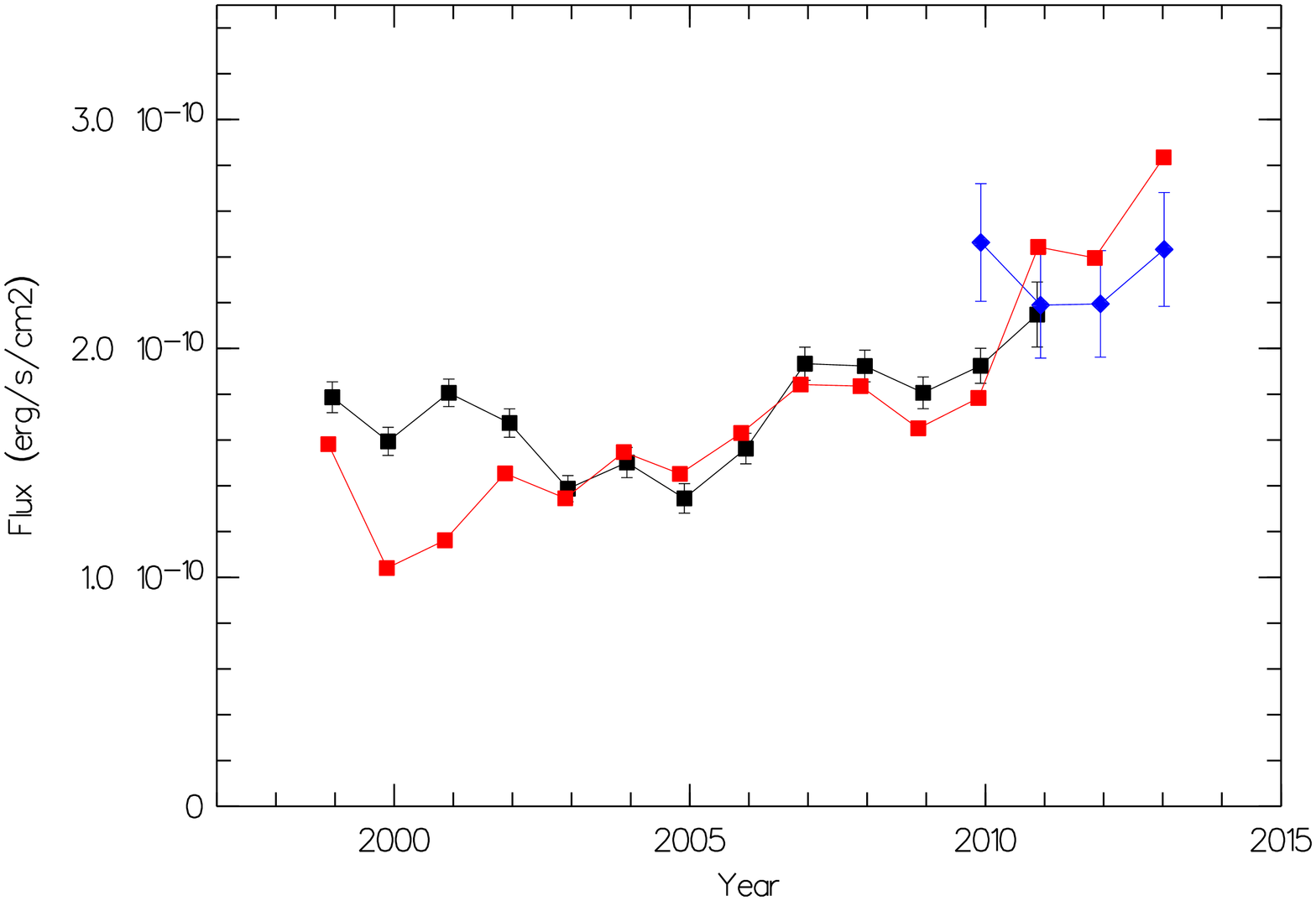}
\end{center}
\vspace*{-0.15in}
% REWORK the caption symbol description for printed version of paper
\caption{{\it Left panel:~} variations of \gc\ of optical/APT flux estimated 
for the Be disk alone against X-ray fluxes from the {\it RXTE} 
All Sky Monitor program over 9 years. 
{\it Right panel:~} running annualized flux averages over 15 years for 
the \gc\ disk (optical, scaled to the X-ray flux range from the relation 
determined in the left panel) and X-ray data with error bars 
(from {\it RXTE/ASM} 
and, for 2010-2013, from the {\it Monitor of X-ray Images} instrument 
on board the International Space Station).  
APT data are open symbols (in red for on-line version). 
% with error bars smaller than the symbol size.  
After \citet{MLS15}.}
\end{figure}

  The red tinge of the 70-day cycle signal indicates that unlike 
the gray 1.21 day signal the long cycles originate from a structure
cooler than the Be star in the \gc\ system. This can only be its
circumstellar disk. These facts argue that the disk is somehow {\it 
associated with} the generation of most or all of the hard X-ray flux.

 SB06 have shown, albeit from optical photometry over a single season, 
that a similar red-tinged cycle may exist for HD\,110432. 
It is a longer timescale variability, $\sim$130 days, 
than the cycles associated with \gc.\footnote{ It can be noted that 
$\sim$70 day cycles have also been found in the B0.3\,IV star $\delta$\,Sco. 
While the phenomenology of these optical cycles has some similarities to 
those discussed here, there are important dissimilarities as well 
\citep{Jones13}. First, the amplitudes are frequently about five 
times as large as those present the in the \gc\ record. 
Second, the color-brightness phasing can vary widely from cycle
to cycle, i.e., from a neutral to correlated to anti-correlated relation. 
The latter is a distinctly different behavior than the consistently 
phased relations in the \gc\ light curve. Third, Jones et al. suggest that 
these variations are caused by the Be star periodically injecting mass
to the disk, thereby inducing structural changes. However, no evidence
exists that \gc\ injects mass into its disk on any quasi-periodic timescale.}

\subsection{Epochal (long-term) changes in the light curves of \gc\
and HD\,157832.}
\label{epchl}

  The first attempt to search the \gc\ X-ray light curve for variations 
on a timescale of years or longer was made by \citet{Horaguchi94}. These
authors found that even when allowance was made for cross-calibration 
uncertainties among early-generation X-ray satellites, 
``no particular tendency toward a long-term variation" were seen. 
Even so, using annual averages of X-ray data, SLM16 have found several 
year-long minor trends in X-ray fluxes of \gc\ (see Figure\,11b) and optical
fluxes that reflect changes in the disk emission. These variations are
mild, amounting no more than tens of percent over a few years.

%FIGURE12
\begin{figure} [ht!]
\begin{center}
\includegraphics[width=2.5in,angle=90]{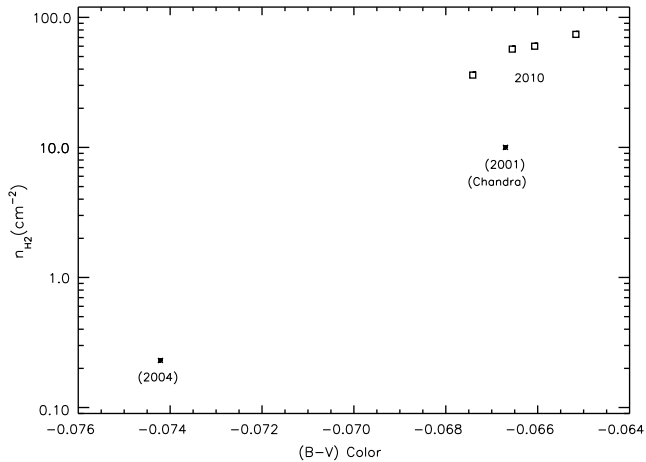}
\hspace*{2.5in}
\vspace*{-0.30in}
\caption{The observed optical $(B-V)$ color magnitude 
against modeled $N_\mathrm{H}$ 
column density (in units of 10$^{22}$\,cm$^{-2}$) for \gc, modeled for {\it XMM}
and {\it Chandra} spectra. The increased $(B-V)$ reddening is caused by a 
continuing build up of the inner Be decretion disk as a result of injections
of matter into the circumstellar environment.  After \citet{Smith12}. }
\end{center}
\end{figure}

   As remarked in the discussion of Figure\,3, variations of 
the {\it soft} X-ray band flux over long timescales are another matter.
First, a factor of three variation was found for \gc\ itself (S04, SLM12) 
and then over 14 years for the \gc\ analog HD\,157832 \citep{LM11}. 
The \gc\ case is particularly edifying, once again, 
because the changes in the soft X-ray
spectrum can be modeled by variations of more than 300$\times$ in the local 
$N_\mathrm{H_b}$ absorption column for $\frac{1}{4}$ of the X-ray site(s). 
According to SLM12, significant changes were also found in the soft fluxes 
during the 40-day interval in mid-2010 over which \gc\ initiated a larger 
than average optical ``Be outburst" in terms of increases in the H$\alpha$ 
emission strength and dimming and reddening of photometric magnitudes. 
In fact, as shown in Figure\,12, the four closely spaced 
{\it XMM} spectra observed in 2010 reinforce a correlation between 
reddening and column density through the soft X-ray attenuation
also seen for the 2001 and 2004 observations. This led SLM12 to conclude 
that foreground gas escaping the Be star during outburst events absorbs
preferentially soft X-rays along our line of sight to the star 
through photoelectric absorption.

\section{The \gc\ analogs.}

The study of the X-ray emission of \gc\ has taken a new direction
with the realization in the last decade when new early-type Be 
stars were discovered with similar X-ray properties (M07, SB06; L06; L07b).
At last, its emission could be compared to those of other stars in 
relevant parameter spaces. These additions revealed characteristics that
turn out to be narrowly distributed not only in the X-ray domain but
also with respect to spectral type, rotation rate, and evolutionary state.  
In addition the presence of H$\alpha$ emission in the spectra of
all these stars indicates the presence of well developed decretion disks.

  Assembling information from the literature, \citet{SLM16}
constructed a catalog of members of
the \gc\ class, given here as Table\,2. The table lists star name, 
spectral type, status of H$\alpha$ Violet and/or Red emission
lobe ({\it s} for single-lobed, {\it d} 
for double-lobed -- this is a proxy for the inclination of the star-disk
plane), equatorial $v\,sin\,i$ from line broadening, 
equivalent width of the H$\alpha$ line, 
binarity (SB1) or blue straggler status (if known),
$V$ magnitude, the energy temperature of the primary plasma ($kT_\mathrm{hot}$),
the X-ray luminosity (taken usually in the range 0.2--10\,keV, 
which we refer to as $L_{\mathrm{x}_{0.2-10}}$), 
and the
citation to the first paper in which the X-ray fluxes were highlighted 
as reminiscent of \gc\ behavior. We discuss each of these stars in order
of their appearance in this table. We have also
added 3XMM\,J190144.5+045914 from the recent 
{\it \xmm}--{\it 2MASS}--{\it GLIMPSE} cross correlation 
catalog \citep[][``N15'']{Nebot15}. 

%TABLE2
\begin{table*}[h!]
\footnotesize
% \scriptsize %smaller font than footnotesize
\centering
\caption{Designated and candidate $\gamma$ Cas stars and some important properties.
(After Smith et al. 2015.)}
\begin{tabular}{l l r c c c r r l}
\hline
 Star Name & Sp. Type & (Lobe) $v$sin\,i & EW$_{H\alpha}$ & SB1?
(Blue  & $V$ mag & $kT_\mathrm{hot}$ & L$_\mathrm{X}$$\times$10$^{-32}$ & Ref. \\
           &             & km\,s$^{-1}$ & (\AA)   & ~~~~~~~~Str.)  &   & (keV)  &
(erg\,s$^{-1}$)      &   \\    
%(10$^{32}$ erg\,s$^{-1}$)      &   \\    
\hline
$\gamma$ Cas & B0.5 IV-Ve & (d) 385 & -34  & SB1  &  2.39 & 12--15.7 & 7--11 & ~(1) \\
HD 110432 & B0.5 IIIe & (d) 350 & -52 & BS & 5.31 & 16-37 & 4.2--5.2 & ~(2) \\
HD 161103 & B0.5 III-Ve & (s) 224 & -31 & -- & 8.69 & 7.4--10 & 4--6: & ~(3) \\
SAO 49725 & B0.5 III-Ve & (s) 234 & -30 & -- & 9.27 & 12.3 & 4--12: & ~(3) \\
HD 119682  & B0.5 & 220  & -30 & BS & 7.91 & 10.4 & 6.2 & ~(4) \\
SS 397   & B1 Ve & --  &  -34 & -- & 11.76 & 6.3--13 & 3.4 & ~(5) \\
NGC\,6649 WL9 & B1-1.5 IIIe & -- & -36 & BS & 11.9  &   ~10 & 5   & ~(5) \\
$^*$XGPS-36 & B1 Ve         & -- & -27 & -- & 14.3: &   ~10 & 3.4 & ~(6) \\
HD 157832 &B1.5 Ve&(d) 217-266 &-25  & -- & 6.6 & 10--11.3 &1.3-3.4 & ~(7) \\
HD 45314    & B0 IVe & 285 & -30 & SB1? & 6.6 & 21 & L$_\mathrm{X}$/L=-6.1               & ~(8) \\
%$^*$HD 93521& O9.5Vp & 390 & -30 & -- & 7.1 & 21 & L$_\mathrm{X}$/L=-5.1& ~(8) \\
$^*$TYC3681-695-1 & B1-2 III/Ve & -- & -31 & --  & 11.36 & -- & 1.4-5.8 &~(9) \\
$^*$2XMMJ ... & ~B0 Ve & -- & -50 & SB?, BS? & 22: & -- & 2.4--3.3 & ~(9) \\
~180816.6-191939 & & &  & & & & & \\
$^*$3XMMJ ... & ~O9e-B3e & -- & -- & --  & --  & -- & 0.4--3.0 &  (10)\\
~190144.5+045914 &    &    &    &    &    &     &    & \\
\hline
\end{tabular}
 \begin{list}{}{}
\item
{\it Notes:}
(1) \citet{White82}, (2) \citet{SB06}, (3) \citet{L06},
(4) \citet{Rak06}, (5) \citet{Motch07}, (6) \citet{Motch10},
(7) \citet{LM11} 
%(8) Rauw et al. 2012 
(8) \citet{Rauw13}, (9) \citet{Nebot13}, (10) \citet{Nebot15}.
\\ 
\vspace*{-0.03in}
\noindent $^{*}$These are ``candidate" objects (see text).
\end{list}
\end{table*}

\subsection{X-ray spectra and optical properties.}

\subsubsection{HD\,110432: the first ``analog."}
 
 As the first discovered and brightest analog excluding
\gc\ itself, this star is the next most studied in the X-ray regime.  
This X-ray source was first found in a HEAO-1 All Sky Survey, and its 
optical counterpart was identified as HD\,110432 by \citet{Cod84}.
The star is probably a member of the Galactic cluster
NGC\,4609. If so, considering its position on the HR Diagram, it is
likely to be a blue straggler \citep{Marco07}. From the presence 
of the Fe XXV/XXVI blend in a {\it BeppoSax} spectrum and the 
afore-mentioned claim of a 14\,ks ``pulse" period, \citet{TO01}
put forward this B0.5\,III star as a binary Be + WD candidate.  However, 
the star was first proposed (SB06) to be a ``\gc\ analog" on the
basis of its hard X-ray spectrum, the undulations of its X-ray light 
curve, and the similarity of its optical spectrum to \gc.~ As noted next,
the X-ray emission properties, while qualitatively similar to \gc,~ 
exhibit somewhat different $kT_\mathrm{hot}$ values and thus gave the 
first rough estimate of the range of properties in this class.

  Two reliable moderate and high resolution X-ray analyses have been 
carried out, first by L07b on the basis of three 
serendipitous {\it XMM/EPIC} observations (limited to wavelengths 
$\le$ 21\,\AA), and second by \citet[][``TSN12'']{TSN12}, who observed 
this star with {\it Chandra}/HTEG. The HTEG yields good signal to noise
ratios over the wavelengths 1.6--16\,\AA.~ These studies were able 
to establish clearly thermality in the plasma by demonstrating the existence 
of three plasma components. However, the values of the respective 
plasma temperatures were different in the two studies ($kT_\mathrm{hot}$ = 
16--37\,keV versus 16---21\,keV, $k$T$_{warm}$ = 3--6\,keV versus 
7--8\,keV, $k$T$_{cool}$ = 0.2---0.7\,keV versus 0.2\,keV). 
Clearly the value of $kT_\mathrm{hot}$
varies substantially over with time for this object, and probably the
temperatures of the warm and cool plasmas as well. 
For both studies the hot plasma emission measure comprised about 80\% 
of the total, which is a similar fraction as we found already for \gc.~ 
TSN12 also obtained and analyzed {\it Suzaku} spectra, which extended the 
energy coverage to 70\,keV. These data disclosed the presence of an 
energy tail that could not be modeled by a 14\,keV $kT_\mathrm{hot}$ plasma
component alone.  Moreover, fitting both spectra over the entire energy
range, these authors found that in addition to the ISM contribution two 
columns were needed to preserve the thermal assumption, one applicable to 
the very hot component and the other to the warm component.
These data discovered a hard energy tail out to 33\,keV
in addition to the hot thermal component. 
The nature of this tail, whether thermal or nonthermal, is so far unresolved.  
TSN12 were also in agreement with L07b that the spectrum of 
this object is likely to be variable over time.

   Flare lifetimes as short as 10--15\,s are observed in the 
light curve of HD\,110432 (L07b), again, suggesting a density of 
$\sim$10$^{14}$ cm$^{-3}$ at the sites of their formation. 
The {\it rif} components of Si\,XIII and S\,XV suggest either high 
densities (if radiative quenching from UV radiation of a nearby star
is unimportant) or, to insure these radiative effects are minimized,  
a distance of less than 2R$_*$ above the Be star for the hot plasma. 
%(or less than 0.4--0.5R$_*$ above a putative nearby WD).

  Another interesting property determined in the spectrum is the pattern 
of elemental abundances. Similar to the case of \gc,~ the [Fe] determined 
from Fe\,L shell ion lines is only slightly subsolar whereas the [Fe] found 
from the K-shell lines is definitely low, e.g., $0.3Z_{\odot}$, though not
quite so low as the \gc\ values in our Table\,1 (L07b). In addition, TSN12 
found that their models underpredicted emission strengths of
Na\,XI, Ne\,X and Fe\,XVII lines, but it is not yet clear that these
signify anomalous abundances.

  A final interesting property of the emission line spectrum is the
large turbulent velocity of 1200\,km\,s$^{-1}$ TSN12 measured the
recombinational Fe\,K lines. This is even larger than the velocities S04, 
L10, and S12 (200--950\,km\,s$^{-1}$) found for lines arising from less 
excited  ions in X-ray spectra of \gc.~
The velocity found for HD\,110432's Fe K-shell lines is interesting 
because it is near the velocities posited in SRC98's flare 
model of exploding parcels.

\subsubsection{HD\,119682.}

   HD\,119682 is a Be star which was first identified as the optical
counterpart of a {\it ROSAT} source \citep[][L07b]{Rak06, SH07}.  
Although the star's position in the 
HR Diagram makes it appear as a young star, 
its kinematic properties and its position in the sky suggests 
that it is a blue straggler member of the Galactic cluster NGC\,5281, and
therefore has an age of 40$\pm{10}$ Myr (e.g., SH07; \citet{Marco09}). 
The spectral type of this star is not yet well agreed upon. Line-blanketed 
synthesis models of its optical spectrum suggest a type of O9.7 \citep{MSH02}, 
while SH07 adopt B0.5. Initial studies 
of {\it Chandra}/HTEG
and {\it XMM}/EPIC by \citet{Rak06}, \citet{SH07}, and \citet{TSN13}
agree that the X-ray continuum is best fit by two plasma components, 
one hot (14.5-15\,keV) and the other cool ($\approx$0.2\,keV). 
They also found an absorption column consistent with the ISM value. 
These studies reported difficulty in finding spectral emission lines. 
However, on the basis of additional {\it Chandra} data, G15 demonstrated 
that the apparent absence of the Fe complex in earlier studies was at least 
partially due to inadequate photon statistics in the extant spectra.
We suspect also that the Fe abundance derived from lines from the K-shell 
ions is subsolar.  \citet{TSN12} found that a single absorption 
column was necessary to fit their data. They argued that this is consistent
with their finding that the Be star-disk system is viewed almost pole-on.

\subsubsection{HD\,161103.}

  HD\,161103 (B0.5III-Ve) was found to have an anomalous X-ray luminosity  
from the {\it ROSAT} All Sky Survey data \citep{Motch97}.  
L06 acquired {\it \xmm/EPIC} spectra to investigate it further.  
The X-ray data suggest 
L$_\mathrm{X_{2-10}}$ $\sim$ 1$\times$10$^{32}$\,ergs\,s$^{-1}$, a hard 
spectral slope, and that the Fe line complex is present. 
Using the same data, L06 and G15 found that two
thermal plasma components are necessary for a good fit to the data. L06 found
$kT_\mathrm{hot}$ =   8.0${\pm 1.0}$ keV and $k$T$_{cool}$ = 0.76${\pm 0.2}$ 
keV. Only a single absorption column was necessary, according to models fitting
the X-ray spectra. It is consistent with the  
ISM column inferred from $E(B-V)$ reddening. As with \gc\ the Fe abundance 
determined from the K-ion lines is probably quite low. 

\subsubsection{SAO\,49725.}

  Like HD\,161103, SAO\,49725 is a B0.5III-Ve star that was 
identified as an early type star with a high X-ray luminosity in 
the {\it ROSAT} All Sky Survey data \citep{Motch97}.  
L06 acquired {\it \xmm/EPIC} spectra to investigate it further.  
The X-ray data suggest 
L$_\mathrm{X_{2-10}}$ 
$\sim$ 1$\times$10$^{32}$\,ergs\,s$^{-1}$, a hard spectral 
slope, and that the Fe line complex are present. 
As for HD\,161103 both L06 and G15 both found that two thermal components
were necessary. L06 found that $kT_\mathrm{hot}$ = 12.8$_{-3.4}^{+7.7}$ keV 
and $k$T$_\mathrm{cool}$ = 0.87$_{-0.25}^{+0.5}$ keV.

G15 found that two thermal components were 
necessary for a good fit to the data, namely 18.86\,keV and 0.82\,keV.
The column density from the X-ray spectrum is consistent with
the ISM column inferred from $E(B-V)$ reddening. Again, as with \gc\ 
the Fe abundance determined from the K-ion lines is probably quite low.

\subsubsection{SS 397; 2MASS\,J18332777-1035243.}

This B0.5Ve star is also known as 2XMM\,J183327.7-103523 and the IR 
source 2MASS\,J18332777-1035243. 
M07, L06, and L07a have identified it as a \gc\ analog. 
These authors found its $kT_\mathrm{hot}$ to be consistent with 9--14\,keV.
The star's L$_\mathrm{X_{0.2-12}}$ is $\sim$4.4$\times$10$^{32}$ ergs\,s$^{-1}$ (N13).
The {\it \xmm} spectrum of SS\,397 also exhibits evidence of thermal Fe lines.
The H$\alpha$ line in the spectrum has an equivalent width of -31\,\AA.~

\subsubsection{NGC\,6649, WL9; 2MASS\,J18332830-1024087. }

This B1-1.5\,IIIe star is the brightest star in the cluster NGC\,6649, 
according to \citet[]{Motch03}. These authors and \citet[]{Marco07} 
judged the star to be a likely blue straggler. 
The X-ray source was first designated as 2XMMJ183328.3-102407 and 
the optical component as either USNO0750-13549725C
or 2MASS\,J18332830-1024087. L06 declared it to be a \gc\ analog based
on {\it \xmm/EPIC} spectra. The N13  catalog reported an X-ray luminosity, 
5$\times$10$^{32}$ ergs$^{-1}$, also consistent with the X-ray emission 
of a \gc\ analog.  L07a found that the spectrum was thermal, based
on the presence of blended Fe\,XXV/Fe\,XXVI lines.
L07a found that the data quality permitted a determination of only a 
single plasma component, consistent with $k$T $\sim$\,11\,keV.

\subsubsection{\it $^*$XGPS-36; GSC2\,S300302371.} 

Also known as 2XMM\,J011559.0+90914 and (optical counterpart) 
GSC2\,S300302371, according to \citet{Motch10}, this star has a spectral 
type near B1Ve and an H$\alpha$ emission EW $\sim$-27\,\AA.~
Assuming a distance of 2--4\,kpc, these authors estimated an X-ray
luminosity L$_\mathrm{X_{0.2-12}}$ $\approx$ 3.4$\times$10$^{32}$ 
ergs\,s$^{-1}$. They also found that {\it \xmm/EPIC} spectra can be fit 
with $k$T $\sim$10\,keV. 

\subsubsection{HD\,157832.}
 
 Of the known \gc\ stars HD\,157832 is the latest spectral type, B1.5\,Ve 
(possibly B2) according to a detailed analysis of its optical spectrum 
by \citet{LM11}. Modeling of {\it \xmm/EPIC} data
suggests a thermal, two-component model, with $kT_\mathrm{hot}$ $\approx$ 
11.25\,keV and $k$T$_{warm}$ $\approx$ 2.3\,keV. The {\it \xmm/EPIC} spectra 
exhibited the presence of the full Fe complex. These authors suggested that 
long-term changes in the X-ray flux is consistent with changes in a
local absorption column, $N_\mathrm{H_b}$.

\subsubsection{HD\,45314.}

  HD\,45314 is another X-ray source with an especially hot plasma
energy temperature of 21${\pm 6}$\,KeV, according to \citep{Rauw13}. 
The full Fe complex presented in the G15 compendium of {\it XMM}/EPIC 
data for this star clearly shows the star's \gc-like nature. 
The short (24\,ks) {\it XMM/EPIC} observation made by these 
authors does not offer good signal-to-noise in order to derive information
about the presence of a secondary X-ray emitting plasma.
Rauw et al. found just marginally different flux levels compared to 
an earlier {\it EXOSAT} observation and thus were unable to judge whether 
the star's flux had actually changed.

  Although we cite these authors' spectral type of B0e\,IV star HD\,45314, 
we also note that \citet{Sota11} assign a spectral type as early as O9:npe.  

\subsubsection{ $^*$TYC3681-695-1; 2MASS\,J01155905+590411.}

This star is also known as 2XMMJ011559.0+590914 and (optical counterpart) 
2MASS01155905+5909141. N13 classified it as a B1-2\,III-Ve star.
This is uncertain because of its large ISM reddening. However, the strong
H$\alpha$ equivalent width (EW $\sim$ -31\,\AA) suggest it is a Be star.
These authors also found that L$_\mathrm{X_{0.2-12}}$  = 
1.4--5.8$\times$10$^{32}$ 
ergs\,s$^{-1}$, which is characteristic of \gc\ analogs. Furthermore, 
the X-ray spectrum is hard \citep{Motch15}. 

\subsubsection{$^*$2XMM\,J180816.6-191939.}

First considered a possible HMXB system by \citet{Motch03}, 
\citet[]{Motch10} obtained additional optical spectra in an attempt to
determine its spectral type, but they could not identify absorption 
lines to determine a spectral type. They determined a rough luminosity 
based on estimated reddening. This luminosity is consistent 
with a main sequence spectral type of B0. They also found that
L$_\mathrm{X_{0.2-12}}$ $\sim$ 3.3$\times$10$^{32}$ ergs\,s$^{-1}$, given an 
assumed distance of 6--7\,kpc. This fact led the authors to advance
this star as a \gc-candidate analog. 
Assuming its estimated spectral type is confirmed by optical spectra, its 
Fe\,XXV/Fe\,XXVI line emissions are also not yet well constrained.

\subsubsection{$^*$3XMM\,J190144.5+045914; 2MASS\,J19014455+0459147.}

N15 cross-correlated the 3XMM, GLIMPSE,
and 2MASS surveys and isolated the source 3XMM\,J190144.5+045914, aka
2MASS\,J19014455+0459147, as a hard X-ray source with a L$_\mathrm{X_{0.2-12}}$
= 0.4--3.0$\times$10$^{32}$ ergs\,s$^{-1}$ and a column density 
of $N_\mathrm{H}$ = 2.7${\pm 0.2}$$\times$10$^{22}$ cm$^{-2}$. 
Its spectral type measured by IR Brackett lines is in 
the vicinity of O9e-B3e, according to the infrared line-spectral type
calibration of \citet{SC01}.
N15 categorically classified this source as a \gc\ analog. 
The spectral type of the star is not accurately defined, nor is the extant 
X-ray spectrum is not well enough exposed to distinguish between a power 
law or a thermal continuum, nor does it show evidence of FeK lines.
However, the XMMFITCAT database \citep{Corral14} reports satisfactory
thermal fits with $k$T $\approx$ 9.1${\pm 3}$\,keV and $N_\mathrm{H}$ = 
3${\pm 0.5}$$\times$10$^{22}$\,cm$^{-2}$. 
The values of energy temperature, X-ray luminosity, and spectral type 
are consistent with \gc\ analogs.
\\

\section{Summary of properties.}
\label{smmry}

From the characteristics of members of the \gc\ class we
may summarize the properties as they now appear. 
\gc\ stars are first of all classical Be stars with spectral 
types in the range $\approx$O9.7--B1.5 and luminosity class from 
III to V, and current strong H$\alpha$ emission. 

 It is now fair to say that {\it if the X-ray spectrum of a Be
star satisfying the above optical criteria and also indicates the presence 
of thermal plasma, i.e. visible Lyman\,$\alpha$ lines of Fe\,XXV and 
Fe\,XXVI, a continuum slope consistent with an energy 
temperature needed to produce them ($k$T $\gtrsim$10\,keV), then it
can be advanced as a \gc\ star.
The spectrum should also exhibit at least a weak Fe fluorescence 
feature (assuming a detectability EW(FeK$\alpha$) $\sim$-30\,eV), 
and the star's L$_\mathrm{X}$ should be 10$^{32}$--10$^{33}$ ergs\,s$^{-1}$. 
% it should be advanced as a \gc\ analog. 
If the detectability of thermal FeK lines
is relaxed, as must be the case for four stars in our Table\,2, marked with
a ``*" symbol, then we will refer to them as ``candidates." If the lines are 
ultimately detected, the ``candidate" qualifier should be dropped and the star
be promoted to full member of the \gc\ class.} 
A remaining task before us is to determine whether the L$_\mathrm{X}$ criterion 
and spectral hardness are sufficient as well as necessary conditions.

  Of the analog stars in our Table\,2 only HD\,110432, HD\,119682,
HD\,45314, and perhaps HD\,157832 are bright enough to have been examined 
so far at high dispersion and at softer X-ray energies than 1\,keV. 
In these cases it is clear that at least one cooler secondary plasma 
is required to fit the spectrum. Similarly, the existence of mild
variability - especially rapid though variability due to ``flares" 
is probably required. However, one cannot add the requirement that the
class should fulfill certain variability criteria, or they would 
not be discovered from X-ray surveys as they have been recently.
In the event of multiple observations over long timescales one
should insist that any variability be small, much smaller for example 
than in most accreting sources.
Other properties such as 
% an indication of a collisionally ionized X-ray plasma and 
a subsolar Fe abundances from the Fe\,XXV and Fe\,XXVI 
lines may also be common if not universal, but this too must be
investigated.

Determining the physical {\it cause} of the hard X-ray emission is 
another matter. We will argue in $\S$\ref{scnro} that, alone, even 
X-ray observations of superb quality and time coverage are insufficient 
to point to a specific mechanism for the generation of the X-rays.

\section{Suggested scenarios for X-ray generation.}
\label{scnro}

\subsection{Accretion onto a degenerate star or its disk.}

\subsubsection{Consideration of the WD accretion scenario (pro and con).}

 The seminal paper of \citet{White82} was pivotal in focusing the
X-ray community's attention to \gc,~ although not necessarily for 
what one might regard today as prescient reasons. These authors 
highlighted the similarity of the optical spectrum of this star to 
that of the prototypical Be X-ray pulsar system X\,Per. 
Also, the presence of a bright nebula surrounding the star 
\citep{PvdB} suggested to these authors that the
then-putative secondary of the \gc\ system is a supernova remnant and
therefore is currently a neutron star. Alas, the authors found no pulses 
in the X-ray light curve. On the other hand, they discovered no FeK feature
or evidence of thermality in the spectrum either, which they might have 
interpreted as being a result of accretion onto a WD. An interesting 
sidenote is that, from the first, these authors recognized that if the 
X-ray spectrum is thermal it means the plasma has $k$T $\approx$ 12\,keV, 
i.e., the currently accepted value.
In any event, they noted that the instrumental limitations existing at
that time
precluded them from detecting intermediate timescale periodicities.  
Moreover, they did not anticipate the presence of rapidly evolving ``flares."  

  \citet{White82} utilized the Bondi-Hoyle accretion radius 
estimated for a hypothesized secondary in the vicinity
of the Be star and were the first to underscore the difficulty of 
powering a luminosity L$_\mathrm{X}$ $\sim$10$^{33}$ ergs\,s$^{-1}$ by Be 
wind-powered  accretion onto a WD.  Moreover, they emphasized
that such emission was unlike that found for OB stars, which at that time
was linked to a so-called ``coronal model" of \citet{M77}.
Altogether, they understood these arguments to favor the 
accretion NS scenario. Then, as noted repeated claims of periodic 
``pulses" in the light curves of various \gc\ analogs for several
years kept alive this scenario  \citep[e.g.,][]{SH07}. 
However, the failure of subsequent investigations to validate claims of 
pulses has since viscerated this argument. 
As already noted, M86 unravelled the NS argument further with 
their finding that the X-ray spectrum of \gc\ is thermal.
%a finding that has been confirmed
%by several studies of a few \gc\ analogs. 
By contrast Be-NS systems exhibit no or at most weak FeK Lyman\,$\alpha$ 
features, whose presence are indicative of thermality, in their 
spectra (M07, M86, G15). 

  M86 considered key characteristics of flares to argue for
their formation on the surface (or accretion column) of a WD. 
These attributes included the emission measures of the flares as well as 
their high plasma temperatures, optical thinnesses, and short durations.  
In the latter case they assumed their short lifetimes arise from cooling 
in a high density medium. The permitted region in the flare-emitting 
blob's radius is consistent with a WD's radius, and this was used 
in support of the WD accretion hypothesis.

The problem with the latter argument is that it is a necessary but not
sufficient condition. Many individual flares may occur over an extended 
surface area, but this does not mean they are generated one at a time over 
the same contiguous surface, e.g., of a WD. To the contrary, one can argue 
that the occasional confluence of flares, including individual flares
having different colors, indicates it is likely that they are 
distributed over a projected area larger than a WD.

  K98 amplified the WD argument from analysis of moderate dispersion
{\it ASCA} spectra of \gc.~ Their results confirmed that the X-ray light 
curve typically contains ubiquitous short-lived flares while its spectrum 
is hard and contains the Fe\,XXV and Fe\,XXVI emission lines. Moreover, 
these authors took the absence of X-ray pulses to indicate that any existing 
WD cannot be strongly magnetic.

 K98 noted dissimilarities of the X-ray spectrum of \gc\ 
with X-ray coronae in late-type active or T Tauri stars, which exhibit 
a continuum and emission line pattern consistent only with cooler plasma. 
They also contended against the contemporary SRC98 picture as they 
understood it and argued further against any ``coronal" model.
%However, some of their criticism 
%was based on misreadings. In particular, 
However, SRC98 had not argued for a coronal model because they emphasized 
the X-ray flux is generated in a much denser medium.  

 In addition, K98 argued that SRC98 did not consider that the light curves
of dwarf nova binaries such as SS\,Cyg could in principle also exhibit an
UV/X-ray anticorrelation such as that found in Figure\,6. 
However, SS\,Cyg is actually a close low-mass binary ($P_\mathrm{orb}$ = 
6.6 hours), consisting of 
a Roche lobe-overflowing K5\,V star and a WD that hosts an accretion disk. 
The system undergoes irregular UV/EUV outbursts lasting on the order of 
a week. Typically, just before an EUV outburst the X-ray flux decreases 
and its hardness changes as well. An analogous mid-UV outburst occurs 
shortly thereafter. Thus, importantly, the changes in these bands do not 
occur simultaneously. In fact, the asynchrony of changes across the 
energy range is explained well by the accretion disk dissipation model 
for dwarf novae. According to this model when an outburst begins, an
instability results from the accumulation of matter at the outer 
edge of the accretion disk and migrates to its inner edge. 
This causes sequential brightening in one waveband after another. 
None of these properties is shared by the synchronous 
%anti-correlation exhibited in Figure\,\ref{hrsxte} and Figure\,9.
anti-correlation exhibited in Figure\,6 and Figure\,9.

  K98 also argued that because flares can be excited on some active
WDs, their existence on \gc\ does not exclude a Be-WD accretion 
scenario. The best example of WDs with flares in their light curves, 
so-called ``intermediate polars." 
Intermediate polars are binary systems 
consisting of an evolutionarily-driven, expanding main sequence star 
and a magnetic WD.  Mass from the primary is transferred by angular 
momentum conservation to an accreting disk around the WD. Matter is then 
guided by magnetic fields to an accretion column at the disk-star interface. 
Initially falling into this column as blobs, these structures are stretched 
vertically by tidal forces. They shock and emit high energy radiation 
when they eventually brake at high densities, emitting flares. Flares
are observed primarily only at certain orbital phases.  
Studies by \citet{BO97} and \citet{Schwarz05} 
have found that X-ray colors of flares can be quite different 
among various polars. They attribute this difference to the
characteristic blob size of a particular polar. In the case of AM\,Her 
small blobs shock near the top of the accretion column, where they can 
be observed by an external observer as moderately hard ($\sim$3\,keV) 
individual flares. In other cases, such as V1309\,Ori, the blobs are
larger and their infalls survive the fall through most of the column
to the WD's surface. The overlying column absorbs the emitted X-ray 
flux, and all that is observed, if visible at all, are aggregates of
unresolved soft X-ray events. \citep[For a review see][]{Mouchet12}. 
However, notice the near equality of the colors of 
most basal and flare fluxes in \gc\ and HD\,110432 is in direct contrast 
to the diversity of flare colors observed among polars.  In addition, the 
observation of ubiquitous flaring in the \gc\ stars' light curves is in
disagreement with the observed phase-dependent events in polar light curves. 
In summary, the flare characteristics of \gc\ are not a good fit to
the characteristics of polars.

\subsubsection{Orbital-rotational relations.}

\citet{C84} discovered that Be-NS binaries follow a strong correlation 
between rotational and orbital periods, where the latter were determined 
from their X-ray pulse rates. They also showed that this correlation
can be predicted by orbiting particles around a strongly magnetized 
neutron star being balanced between centrifugal and magnetic forces at
points where the corotation and Alfv\'en radii are equal.
The determination of periods for additional 
Be-NS systems over the years has confirmed Mason \& Corbet's results, and thus 
the so-called Corbet relation has been a useful tool in predicting 
one period or the other when only one is known. Building on this
result, \citet{App94} has applied the Corbet relation for Be X-ray 
binaries with shorter pulse periods. His reasoning was that shorter
pulse periods might be expected for Be-WD systems because of the 
larger radii and weaker field strengths of white dwarfs.
Apparao found that two binaries in his sample fell on a line in the 
P$_{orbit}$--P$_{rot}$ plane well to the left (towards shorter orbital 
periods) where most Be-NS systems reside. He argued that the shorter 
orbital periods occur because the degenerate secondary stars have 
weaker fields than NS do, and he took these as possible Be-WD binaries.
\citet{App02} found next that the 204\,day orbital and $\sim$1 day
{\it rotational} periods
of \gc\ fall close to the putative Be-WD relation and therefore that 
systems falling along it can be considered candidate Be-magnetic 
WD systems.  In assessing this hypothesis, one can
point out that no evidence has since been uncovered to confirm that 
Apparao's sample of two are in fact Be-WD systems.  Indeed, in the
intervening years no Be-WD systems have been found that emit hard X-rays.
Also, in a review of the status of the Corbet relation \citet{Knigge11} 
find a similar subpopulation to Apparao's.  However, they judge 
it to be secondary population of Be-NS, not Be-WD, systems. 
It is worth emphasizing again that recent evidence demonstrates that robust 
``pulses" are not present in the X-ray light curves of \gc\ or its analogs.

\subsubsection{Super-soft X-ray sources associated with Be stars?}

  According to the theoretical models of \citet{R01}, $\sim$70\% of all 
Be stars could be formed during late stages of binary evolution as a
result of angular momentum-rich material migrating through the Roche lobe, 
accreting to and spinning up the surface layers of the Be star. If so the
typical secondaries are white dwarfs with $T_{\mathrm eff}$ $\sim$40\,kK. 
The fact that such secondaries are so much fainter than B stars, even
in the UV, is one reason for the general failure to find such systems.
If so, one asks: what would they look like in the X-ray domain?  
Might Be-WD systems be, for example, a subclass of so-called Super-Soft X-ray 
Sources (SSS)?  Could these be related to or confused with \gc\ stars? 

 SSS are thought to represent a broad class of very soft X-ray emitters 
that are detected at energies only below 1\,keV \citep{Kahabka97}. 
They occur mainly in binaries, including subclasses of novae and 
polars, but they can be found also among evolved single stars (nuclei of 
Planetary Nebulae and cooling NS). Under the right conditions nuclear
burning occurs when matter from an evolving primary falls and accumulates 
onto the WD surface. Upon thermalization of the released potential energy 
a ``super-soft" ($k$T $<$ 100\,eV) black body spectrum-like is emitted. 
The X-ray flux is strongly attenuated by a dense accretion disk around 
the WD. Although this makes it difficult to determine the intrinsic 
L$_\mathrm{X}$ of the source accurately, typical estimates lie in the range 
of 10$^{36}$--10$^{38}$\,ergs\,s$^{-1}$. Recently survey work has 
uncovered one SSS associated with an early-type star in each of the 
Magellanic Clouds, detailed as follows.

\citet{Kahabka06} discovered the source XMMU\,J052016.0-692505 in
the {\it \xmm/EPIC} public archive and identified its optical counterpart
as the Large Magellanic Cloud star LMCV2135. Optical spectroscopy revealed
that this to be an early Be giant star that hosts an extensive disk.  
Moreover, UV filter magnitudes from the {\it \xmm} Optical Monitor 
are consistent with a photometic spectral type of a B0-B3\,III star and
a mass near $16\,M_{\odot}$.  At the time of this discovery a Be star 
with a super-soft X-ray spectrum was a novelty. 
Thus, they concluded that this SSS Be system is unique and that
its X-rays resulted from thermal runaway of matter steadily accreted 
onto the surface of a WD. If so, it has become the first 
known Be-WD system. The authors bolstered their argument by 
showing that the evolutionary-determined ages and the masses estimated 
for the binary components are consistent with evolutionary scenarios 
predicted by \citet{R01} models for high mass binaries.
In sum, these authors argued that SSS objects are a logical channel for
Be stars that have been spun up by binary mass transfer. Indeed the
spinning up of Be stars in close binaries appears to be a common 
occurrence, according to a study of Be stars observed in open clusters
\citep{McSw05}.

Recently \citet{Sturm12} conducted a survey for SSS in the Small
Magellanic Cloud and discovered the SSS object XMMU\,J010147.5-715550
and determined its optical counterpart to be  AzV\,281, an 
O7\,IIIe--B0\,Ie star. 
Using arguments similar to those of \citet{Kahabka06}, they concluded
that this is another Be-WD system.
In addition, X-ray monitoring of the object disclosed the unique 
circumstance that its soft band (0.2-1.0\,keV) 
flux decreased by a factor of $\gtrsim$40$\times$ in roughly a year.

Neither of these authors suggested a connection between \gc\ and
SSS objects. Indeed \citet{Kahabka06} drew a contrast between the two
on the basis of the very softness of SSS star spectra. 
In addition, the principal sculpting agent of X-ray spectra of SSS Be 
star binaries is the dense accretion disk surrounding the secondary star; 
there is no apparent role for the decretion disk of the Be star.  
Altogether, there is scant if any evidence for a connection between the 
two groups of objects.

 To conclude this section, the NS accretion hypothesis has lost its
early luster in favor of the still-surviving WD one. The surviving 
appeal of the WD picture lies in the similarity of its spectrum to 
certain types of active white dwarfs (especially intermediate polars) 
% the nearly circularity of the \gc\ orbital system, 
and the absence of pulses in either \gc\ itself or its analogs.
A lingering problem with the WD accretion scenario, as noted 
by \citet{White82}, has been the question of whether the
powering of the observed L$_\mathrm{X}$ $\sim$10$^{33}$ ergs\,s$^{-1}$ can
be accomplished from matter accreted from a WD in a \gc-like orbit. 
Estimates in the recent X-ray binary literature \citep[e.g.,][]{Sturm12} 
suggest a range of 10$^{29}$--10$^{33}$ ergs\,s$^{-1}$. The upper limit to this 
range assumes rapid expulsion of matter in the plane of the disk by 
a discrete ejection - since it must be orders of magnitude above a
wind mass loss rate. 
Part of the issue has been the uncertainty of the outflow velocity that
discrete ejections have. This is important because the mass loss and hence 
putative accretion-powered L$_\mathrm{X}$ 
increases as 1/$v$$_{outflow}^3$, according to
the Bondi-Hoyle accretion formalism. Probably a more relevant question is how 
efficiently the outer edge of the disk of a \gc\ star is truncated by the
tidal interactions with a binary companion (see $\S$\ref{binry})
and whether discrete mass ejections can cross such a
barrier rapidly enough before dispersing to the ISM or 
returning to the Be star. Although this issue is not yet decisively
resolved, so far the evidence does not favor rapid crossings, if 
such crossings occur at all (e.g., MLS15). In any case, the optical 
variations resulting from a strong Be outburst are seldom events. 
They do not occur in a continuously-repeating or sustained manner,
e.g., \citep{Pollmann14, HS12}, such as would be required to power the 
full observed X-ray luminosity of \gc.

\subsection{ The Magnetic Star-disk hypothesis.}

\subsubsection{ Description.}
\label{msddesc}

Much of the advocacy for the picture of accretion onto a degenerate 
companion has been made by attempting to tie the \gc\ case to 
other classes of X-ray Be binary systems
for which the accretion model has been so successfully applied.
In this section we argue that overall this is insufficient because
the \gc\ stars comprise a unique class, one which
associates X-ray properties to the immediate environment of the Be star
rather than to a binary companion. With this as context, we introduce
the so-called magnetic star-disk interaction picture. We preface
our discussion by pointing out that this is a hypothesis built on the 
empirical data alone. It is supported by a few conjectures, as we will
see,  and it is not yet capable of making predictions.

 The original hypothesis drew its support from correlations of
properties of X-ray and UV/optical diagnostics. 
The first multi-wavelength correlated event was an X-ray flare in the
{\it Copernicus} light curve of \gc~ reported to \citet{Peters82} in
1978. This coincided with short-lived emissions in the UV resonance
lines of Mg\,II and Si\,IV and also an H$\alpha$ emisson ``flare" 
\citep{SS78}. 
However, this was a single event.
The best contnuing array of events
was the simultaneous UV/X-ray campaign in 1996.  The X-ray/UV 
correlations reproduced in Figures\,6--8 suggested a picture of short-lived 
corotating cloud complexes ejected from active sites on the Be star
(see $\S$\ref{hdbas}).  Next, the discovery 
that the 50--91 day cycles are present in both the optical and
X-ray region (Figure\,10), together with their amplitudes being 
about larger in the red than the blue, suggested that the cycles
originate in the cool Be decretion disk (RSH02).
As noted, MLS15 showed that this initial finding can be broadened to 
demonstrate that optical and X-ray variations of even a year or longer 
are well correlated, with zero delay and with the same scaling
factor applying to this broadened range.
Importantly, the zero delay is at variance with the several year
delay expected in the accretion scenario between the optical outburst, 
marking ejection of matter from the Be star and the corresponding
X-ray surge due to accretion onto the remote companion.  }
Since the optical variations are well understood to arise from the
development of the inner regions of the Be decretion disk, a 
reasonable conclusion is that the reservoir responsible for hard X-ray 
energy generation is associated with this structure.  {\it From this
inference the concept of a magnetic interaction between the Be star 
and its disk was born, a picture that is sure to evolve further with 
new discoveries.}

 In brief, the scenario is as follows. By hypothesis, magnetic field lines 
extend from points on the Be star's surface, at first in the form of 
small-scale loops. Consider next that the angular rotation rate at the star's
surface is greater than the Keplerian rate of layers within the inner decretion 
disk.  Magnetic field lines embedded in the disk interact with the stellar 
field lines, at first gently. They quickly become entangled and stretched 
because the two fields are anchored to their respective sources. 
The stretching continues until eventually the lines sever and reconnect. 
They attempt to come to equilibrium by springing
back to a lower magnetic potential and in so doing accelerate particles
embedded in them (see \citet{Zirker12} for a readable discussion of 
these processes in the solar context). In the \gc\ environment, they are 
expected to be accelerated in various directions, including toward the 
Be star and probably also the disk.  
Figure\,13 depicts this action as a cartoon.

\vspace*{0.15in}
% FIGURE13
\begin{figure}[ht!]
\label{asm}
\begin{center}
\includegraphics*[width=10.0cm,angle=00]{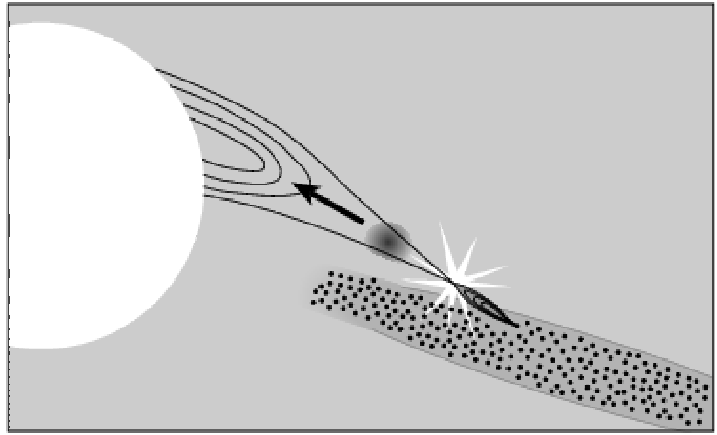}
\end{center}
\vspace*{-0.15in}
\caption{ Depiction of the interaction between putative magnetic loop
systems from \gc\ and its decretion disk as seen from the equatorial 
plane. The cartoon shows the acceleration of matter following reconnection 
of field lines after their initial entanglement and severing.  
Particle accelerations may be directed toward the star or the disk. 
After \citet{RS00}.}
\end{figure}

  By hypothesis the field reconnections occur as discrete events,
propelling some of the particles toward the star 
at a few thousand kilometers per second.
They impact points on the star, 
and thermalize there, causing the impacted gas parcels to expand explosively. 
We interpet these events as X-ray flares. In their analysis of
flare dynamics SRC98 showed that the emission measures and durations
of the overwhelming fraction of flares expand adiabatically, thereby
maintaining their temperature during this expansion phase. 
Note that SRC98 did not have to specify the source of the high energy.

  RS00 took the next step by following the theory of \citet{Wheatland95} and
constructing thick target bombardment simulations of electron streams having
an energy temperature $k$T of 200\,keV onto a boundary having parameters of the
photosphere of a B star. With these parameters they found a braking depth
of the stream that corresponds to a gas density of nearly 10$^{14}$ cm$^{-3}$ 
and a characteristic energy temperature of $k$T $\sim$ 10\,keV. 
These are precisely the parameters that SRC98 had derived from
their observations of flares for \gc.~ 
To put some numbers into this concept, consider the case of an 
exemplar flare with an emission measure of 10$^{55}$\,cm$^{-3}$.
The flare is formed
in an exploding surface parcel with a radius of 5$\times$10$^{3}$\,km. 
RS00's high energy (200 keV) electron beam would yield an energy of 
$\approx$4$\times$10$^{32}$ ergs\,s$^{-1}$ across the parcel's surface. 
This is sufficient to power the flare for about a second. Note that 
this is a lower limit because it neglects the high energy tail of the 
electron distribution that RS00 thought might also be necessary in the
real case. It also neglects  the virialization of the initial energy of 
the input beam. 
Additionally, the value of $kT_\mathrm{hot}$ now generally agreed upon for
\gc\ is at least 50\% higher than the value of 100\,MK (8\,keV) RS00 used.
Thus, we know the energy numbers of RS00 are low.

This work also demonstrated 
that an electron beam with a monoenergetic distribution leads to a flare
parcel temperature slightly in excess of the temperature of a large 
magnetically-constrained volume that emits background (basal) flux. 
If a power law rather than a monoenergetic distribution is assumed for
the injected beam, the inequality is reversed: T$_{flare}$ $<$ T$_{basal}$. 
Both inequalities have been observed, sometimes within tens of
minutes of one another in the X-ray light curves of both \gc\ and 
HD\,110432 (SRC98; SLM12). We remind the reader that the
existence of magnetically constrained volumes leading to this basal emission 
is suggested by the existence of corotating clouds.

  If particle streams are directed to the star from instabilities near 
the inner disk, one can seek to observe them. SR99 reported such evidence 
in their analysis of spectral time series of 1045 {\it GHRS} spectra in
the March 1996 campaign on \gc.~ These moderate resolution spectra had 
a very high signal to noise ratio, thereby allowing analyses of variations
of even weak spectral lines. 
The authors constructed a computer program that searched through the 
time series for similar responses in absorption features separated by 
wavelength separations corresponding to several pairs of `related" 
spectral lines. Related line pairs could be members of a multiplet 
or they might arise from two ions having similar ionization potentials.
%and they also had to have similar strengths. 
Cross-correlations were carried out for fluctuations of all
spectra in the time series and across a broad velocity domain.
The result was a velocity-time map in which ``excess" cross-correlations
could be identified and compared with correlations for fluctuations 
at random wavelength spacings, which served as controls. 
The results showed two interesting patterns. First, the excess 
cross-correlations occurred, with one exception, only for narrow bands 
centered at positive velocities. The exception was the strong correlation 
discovered at a range of negative velocities for the Si\,IV resonance 
line doublet -- this is expected for these well-known diagnostics of 
an outflowing hot-star wind. The second pattern was that the best 
cross-correlations were found at positive velocities that increased 
according to the excitation ion state of the lines. For example, for pairs 
of Fe\,V and Si\,IV lines corresponding to a hot gas, excess correlations 
occurred preferentially for velocities $\gtrsim$+2000\,km\,s$^{-1}$ while 
for a pair of low excitation Ni\,II and S\,III lines excess correlations 
occurred for positive velocities of only $\sim$+200\,km\,s$^{-1}$. 
All told, this analysis argued that the excess correlations were real and 
constitute evidence of high velocity matter expelled toward the Be star. 
Moreover, velocities of $>$2000\,km$^{-1}$ are much greater than the 
escape velocity at the surface of a B0.5\,IV star. 
This fact rules out an interpretation that the features arise 
as returning matter from an earlier failed ejection. The only 
reasonable scenario is that the matter had been accelerated 
toward the star by some intervening structure in our line of sight.

The concept of a disk dynamo was suggested on the basis of the correlation
between the optical and X-ray $\sim$70 day cycles. To our knowledge
this is the first suggestion for a dynamo for any stellar disk.
According to this idea, a dynamo is set up in any disk in which the following
conditions are met: 
(1) a seed magnetic field is present within the disk, (2) the angular 
velocity of the Kepler disk around the star decreases with increasing 
distance, and (3) no external field intrudes into the disk and disrupts its
field.
In such disks particle motions will be subject to a magnetorotational 
instability (MRI), in which a a positive feedback loop is set up that 
increases the disk seed field and gas turbulence. 
Orbiting gas parcels respond in a dynamo cell by alternately moving inwards 
and outwards during a cycle (as specific angular momentum does the opposite).
%  -- that is, they move inwards and outwards from the star. 
During the phase when particles approach their innermost radius, the gas is 
compressed. 
This increases the local emission measure, which goes as electron density 
squared, thereby adding to the luminosity of the star-disk complex. 
The upshot is that particles and field lines in the inner disk experience 
cycles, and as they do the instances of entanglement with stellar field 
lines are modulated as well. This produces cyclic variations in 
the numbers of accelerated particle streams and generation of flares. 

  In considering the conditions in which a disk dynamo might operate, a
Keplerian rotation law fulfills the orbital velocity-distance requirement.
In practice, seed fields can be assumed to exist already (SHV06). 
The requirement that internal disk fields are not distorted by external 
ones cannot yet be evaluated. In evaluating the practicality of this 
mechanism, the overriding question is not so much whether dynamo actions 
exist in the disk of \gc\ but rather whether the instabilities of individual 
cells become organized into a coherent, disk-wide dynamo. 
Otherwise, local chaotic motions result without a clearly observable 
global effect. Another issue is that the mechanism is {\it ad hoc} - it 
is constructed so far entirely from the phenomenology. 
However, one clear inference is that energy from magnetic stresses 
will be periodically exhausted in the inner disk, thereby  mediating 
the production of accelerated particles. In addition, the picture 
suggests injections of streams distributed at more or less random times 
such that particles will be accelerated towards all possible longitudes 
on the Be star surface. Otherwise, quantitative predictions are not yet 
possible. In particular, the natural period of dynamo cycles is the 
orbital period of the inner disk, which in this case is no more than a 
few days, not 70 days. 
In sum, this hypothesis needs
confirmation, including both continued theoretical and laboratory work.
It is promising, but it may or may not be the correct instability 
underlying the observed disk variations.  Ultimately, it must be
understood why this mechanism would work only for some Be stars.

 One can also ask how the magnetic fields on \gc\ might be generated 
because we know that organized low-order fields do not exist on the
star's surface: if they did, their spectral UV resonance lines would 
show a telltale, regularly changing pattern of emission and absorption 
features due to clouds frozen into the stellar dipole field. 
Recent magnetic surveys 
of Be stars have disclosed that organized, low-order magnetic fields 
are not detected (Wade et al. 2015). MLS16 have pointed out that 
of the 18 Be stars for which rotational/disk obliquities are known 
from Long Baseline Optical Interferometry, the angular rotation rates
of \gc\ and HD\,110432 are the ones most likely to be close to
their critical limits. The effects of destabilizing hydrostatic
equilibrium are the greatest in the outer equatorial envelopes
of such stars. In the fastest rotating stars the otherwise thin convective
envelopes of these stars 
(which owe their existence to large opacities from Fe peak ions) 
can increase dramatically \citep{Maeder08}. Therefore, particularly 
in the case of differential rotation, local dynamos in the stellar 
envelope could develop that would spawn disorganized, nonpermanent 
fields -- see \citet{Cant2009, Cant2011}.
Such fields could well elude spectropolarimetric surveys 
such as \citet{KS13}, even though 
their field lines could reach out to the inner regions of the Be disk.
While this last argument is speculation, we note from variations 
in the waveform of the 1.2-day photometric signature that the surface 
``patch" corresponding to an embedded structure on the surface of \gc\ 
has been observed to change on a timescale of two years (HS12). 
During the same time interval the amplitude of the photometric
signature diminished by factor of 3--4.  
The amplitudes and waveforms were constant to within errors during the
few years before and after this interval. However, by the 2013--2014
season the 1.2 day signature was only marginally present in only 
one of two-filter APT light curves. By the following season it had
disappeared altogether \citep{Henry2015}.  These observations are 
consistent with a constantly-evolving, chaotic, but active field 
anchored not far under the star's surface, again following the
Cantiello \& Braithwaite prescription.

  We have no evidence as to how, or indeed whether, the \gc\ stars have 
been ``spun up." It may be due to a single massive star's having an
especially large angular momentum, or it may be that the spin up
occurs from a prior mass transfer thanks to a evolved, now passive
degenerate companion. One argument favoring the latter hypothesis
is that three of the \gc\ analogs are probably blue stragglers (Table\,2).
Could most or even all of them be blue stragglers?
This is a subject for much more research.

\subsubsection{Observational underpinnings of the interaction hypothesis.}
\label{undrpn}

  We recapitulate this section by listing the observational underpinnings
of the magnetic star-disk interaction picture. Most of these have been
introduced already. We summarize them in rough order of their chronology of
the discovery that the star is a hard X-ray source.

\begin{enumerate}

\item{X-ray correlations/anticorrelations are evident in the 
intermediate-timescale
variations of UV diagnostics.
The key point is that in the Be/disk/binary complex only the Be star 
emits enough UV continuum and line fluxes for variations to be easily 
observed. The time coverage of the simultaneous 1996 {\it RXTE/HST} 
observations was too long (Figs.\,5--7), and both the number of up/down flux
excursions and the pattern too self-consistent, to have occurred by chance. 
One concludes that the X-rays are created close to the Be star. } 

\item{ X-ray flares of \gc\ and HD\,110432 
must be emitted in a high density ($\gtrsim$10$^{14}$ cm$^{-3}$) environment.
This limit is set by their short lifetimes ($\S$\ref{flrprop}) and the ratios 
of spectral emission lines arising from Fe\,L-shell ions ($\S$\,\ref{rifdns}). 
Only the photospheres of the two binary components could be the site
of these flares. From the consideration of Bullet 1 one is left with
the Be photosphere as the site of flare production.
}

\item {An attenuation of soft X-rays was observed that coincided precisely
with the optical Be outburst of \gc\ in 2010 ($\S$\ref{epchl} and Figure\,3).
As this outburst proceeded over the 40 days of the {\it \xmm}
observations, the spectra showed
progressive attenuations of the soft X-ray flux, thereby requiring an
increased local absorption column $N_\mathrm{H_b}$ to fit the spectrum. 
At visible wavelengths, an optical outburst generally unfolds as a
brightening and reddening of the flux from the star-disk complex. Again,
Figure\,12 showed the relation between $(B-V$) color and the determined 
$N_\mathrm{H_b}$ column. The figure overplots the same parameters
during the 2001 and 2004 X-ray observations, so we can see
a common relation between these variables during different these times.}
Because the emission of the hot primary plasma dominates at both short 
and long X-ray wavelengths, the column obscuring the soft X-rays also 
obscures hot X-ray emitting sites in the background. Since the column 
represents an injection of matter from the Be star to its
circumstellar environment, this pattern indicates that the 
hot plasma sources are situated at the foot of the column, i.e., just
above or even at the surface of the Be star.

\item {No delays are observed in integrated X-ray variability 
relative to optical variations of \gc.~
The MLS15 extension of the RSH02 and SHV06 work that the scalings 
of correlated X-ray and optical variations over a timescale from tens of 
days to a year or so is important because it demonstrates that only 
a small (dense inner) region of the $V$-emitting decretion disk is
involved in the optical/X-ray correlations.
MLS15 also investigated the possibility that discrete Be ejections could 
transit across the binary system and accrete onto a degenerate companion 
such as a WD. 
% thereby producing X-rays. 
However, they recognized that
such ejecta would take a finite time to travel past the outer 
truncation and Roche lobe radii before falling onto the 
putative WD. These authors found no delay  in the X-ray variations
behind the optical ones.  Using Monte Carlo techniques to noise to their
{\it APT} and {\it ASM} data, the authors estimated errors on this
null result of one month.
In contrast, they pointed out that two well known X-ray Be binary systems with 
comparable orbital separations are {\it observed} to show a delay of a few
years of X-ray responses following optical outbursts of the Be primaries. 
On this basis, the authors argued that any X-ray generation 
mechanism requiring accretion onto a degenerate secondary in the
orbit known for the \gc\ companion is inadmissible.
}

\item{Long-cycle variations are mediated by conditions in the inner 
part of the Be disk.
The good correlation of X-ray and optical long cycles noted by SHV06
and MLS15 demonstrates that the circumstellar Be disk is 
somehow {\it associated} with X-ray generation. According to the star-disk 
interaction scenario, the disk provides a storage medium for pent up
magnetic energy that is released at semi-periodic intervals. 
In this sense it represents an intermediate link in a chain of 
processes leading to particle acceleration and ultimately X-rays.
% emission in flares and corotating, confined cavities. 
 }

\end{enumerate}

\section{Summary \& Conclusions.}

\subsection{Summary.}

   In this paper we have reviewed the study of X-ray properties of
\gc\ to date. Work over the last several years has clarified that 
this star is actually the prototype of a class of peculiar X-ray emitters
among those associated with massive hot stars. 
To the extent that observations of fainter stars can unveil them, 
these studies reveal remarkably similar properties in the X-ray, UV,
and optical domain. Part of this 
review has been an evaluation of candidate mechanisms for the emission 
of the hard X-rays, including one in which accretion onto a degenerate 
companion, with its deep gravitational well, could release X-rays. 

The Be-NS scenario, favored for a decade or more after \citet{White82}, 
became disfavored following a consensus that the X-ray spectrum can be 
fit not by a sum of black body spectrum with an additional nonthermal 
(Compton) tail 
but rather by the combined emission of multi-component, optically 
thin thermal plasmas. In particular, emission lines from the FeK complex at 
$\ltsim$2\,\AA\ and others arising from light-element ions 
in the soft X-ray regime are present in the spectra of \gc\ stars. 
In addition, the 
X-ray light curves of Be-NS systems are dominated either by 
strictly periodic pulses, as in highly magnetized Be X-ray pulsar 
systems, or semi-regular outbursts tied either to the orbital period 
or unpredictable optical Be star events. 

It took some additional time for
a consensus (albeit perhaps still not unanimous) to arise that the Be-WD 
systems cannot fit the observed multi-band phenomenology. It is still 
the case that the efficiencies of mass transfer and thermalization must 
be exquisitely fine-tuned to match the X-ray luminosities of these systems, 
particularly with what we now understand the mass loss rates of \gc\ and
other classical Be stars to be.  Furthermore, searches for Be-WD systems 
have consistently come up either as altogether negative \citep{GN09} 
or as hypothetical suggestions of special cases, such as super-soft X-ray 
systems, that do not mimic the X-ray characteristics of \gc\ stars. 
In addition, the simultaneous array of multi-component 
thermal plasmas with ubiquitous, ``gray" flaring are attributes 
that are rare in active WD systems, if present at all.
 
  The strongest arguments for the magnetic star-disk interaction picture
- which is undoubtedly incomplete at some level - are corroborative,
rather than exclusionary. As detailed in $\S$\ref{undrpn} these
arguments pertain to X-ray variations and their correlations on a 
wide variety of timescales with UV or optical diagnostics. Here it
should be understood that these diagnostics are to be expected to be 
formed over the surface of a $\sim$B0.5 star.

In the following we will pose the most relevant questions that can 
at last be asked about this class of stars. We follow with a few 
desiderata on what needs yet to be done. We do this with trepidation
because the history of this subject has been that progress has often
come from very unexpected sources.

\subsection{Questions, resolved or nearly so.}

\noindent {\it Can we easily recognize a \gc\ star?}

In principle, yes. Given that optical spectra show that an early B star is
a Be star neither embedded in a star-forming region or a component 
of an interacting binary, the work of L07a and G15 have strongly 
suggested that if its X-ray spectrum exhibits the full FeK line complex 
and a nearby continuum slope that supports the energy temperatures 
(i.e., $\gtrsim$7\,keV) needed to maintain the strength of these 
lines as well as an L$_\mathrm{X}$ of 10$^{32}$-10$^{33}$ ergs\,s$^{-1}$,
the star may be reasonably advanced as a \gc\ analog.  
These circumstances favor a wholesale discovery that has been so
successful in the last few years in expanding the class,
particularly thanks to cross-correlations of {\it \xmm/EPIC} 
surveys with optical or even infrared catalogs. The sample as 
now recognized is probably complete out to a distance of $\sim$2\,kpc, 
at least for the area covered by the {\it \xmm} surveys. The coverage can
be furthered by continued cross-correlation of hard X-ray and infrared
catalogs in the Galactic plane.
\\

\noindent {\it Do we have a large enough sample of \gc\ stars yet 
to constrain the range of their X-ray characteristics?}

From the sample of stars in Table\,2 we now have a well defined
parameter space which helps to constrain the X-ray generation mechanism 
resides. Some areas where improvement could be required are covered by the
following questions: can a late-type Oe
star assume the X-ray properties of \gc\ stars during the much
briefer times their decretion disks are maintained? Is there truly
no overlap at B2 with the Bp stars, where both ordered and disordered
fields could exist simultaneously on a star's surface?  
Also, what is the lowest energy temperature an X-ray plasma hosted 
by in a \gc\ star may have?  At the moment it appears to be about 
6--7\,keV. The physical conditions setting this limit will be relevant 
to understanding the geometry and magnetic topology of the X-ray 
producing environment, possibly including the status of the Be inner disk.

\subsection{Desiderata.}

A primary issue to resolve is the role of binarity in the 
setting up of the X-ray generation mechanism in  \gc\ stars. 
This question comes up first because 
of the known binarity of \gc\ itself and second because of the
likelihood that at least three analogs are old blue stragglers in Galactic
clusters. An answer to this question will disclose critical details 
about their past evolution, their current intrabinary interactions and, 
most especially, whether binarity is an essential element 
that causes the Be star to spin up to criticality.
Thus, we recommend radial velocity monitoring in order to define the 
binary status of \gc\ stars -- the only known \gc\ 
binary so far being \gc\ itself.
The questions to be asked are:
what are the ranges in semimajor axes and orbital eccentricities?

We have speculated that an early Be star's rotating
at near the critical velocity is important in setting up local
envelope dynamos that create unstable local magnetic fields. 
There are three approaches to test this idea. The first is
to explore the efficacy of diffusing excess angular momentum
deposited on the star's surface by transfers of matter from a
putative companion. This question can be addressed theoretically 
or by an empirical study of angular rates inside a B star determined 
from rotational splitting of nonradial pulsation modes. 
The second approach is to address the critical rotation question
observationally by confirming that a near-critical rotation rate 
is important to the creation of the needed surface conditions
for the mechanism to develop, as suggested by MLS15. Improvements in 
LBOI instrumentation (or increases in the total integration time) would help
refine estimates in the obliquity angles of the rotation planes of some
of the brighter analogs.

The third approach is to survey binarity 
of a bright subsample of \gc\ analog stars from long-term
studies of their radial velocity curves. For example, a
single, compelling nondetection of a nonvariable RV curve of 
an \gc\ suspected to viewed edge-on (from the presence of double-peaked 
emission profile as well as absorption ``shell" lines) would go a 
long way to refuting the question of binarity. 

 The report by \citet{Henry2015} that the rotational signature in
the light curve of \gc\ has disappeared now 
invites the expectation that it will return.
Further APT observations will tell us how long it takes to reappear, 
grow to full strength and indeed whether the patch responsible for it 
will reemerge on the surface at the same place (longitude).

  A remaining important question is whether the presence of a stable 
Be decretion disk, as measured by some minimum EW of H$\alpha$ 
emission, is indeed necessary for the production of hard X-rays.
Over the coming period, whether it be measured in years or a few 
decades, astronomers should be able to monitor the H$\alpha$ variability
enough to catch the transition of one of the \gc\ stars in Table\,2 as it
slips into a non-Be phase. When this happens it will be of great interest 
to verify whether the X-ray emission quickly subsides and, assuming this 
to be so, to examine how the specific spectroscopic X-ray diagnostics 
change during the transition to the quiet phase.

Our final rhetorical question is whether a search for faint reflection 
and emission nebulae similar to IC\,59 and IC\,63, which are spatially 
correlated with \gc\ \citep{PvdB}, might reveal nebulosities in the 
disk planes of other \gc\ stars. One wonders:
are such nebulosities common around \gc\ stars but absent from other 
active Be stars?  In addition, these authors suggested that a 
light echo could be observed in such nebulae some years after
a large optical outburst of the Be star (as would be expected 
now after three years in the case of \gc.~

All these investigations are aimed at answering the key question: how does 
the hard X-ray emission from \gc\ star form?  Clearly the development of the 
``\gc\ phenomenon" requires the emergence of a complicated chain of processes. 
We have identified a number of observational disciplines and studies that are
necessary for this development. They include
X-ray and optical spectroscopy and photometry, Long Baseline Optical 
Interferometry, as well as a  comparison of characteristics of \gc\ 
analogs to accretion-related phenomena known to occur on other evolved 
binary systems. 
However, these disciplines alone are unlikely to reveal the {\it linkages} 
in the chain of processes that produce hard X-rays, and so new ones will be 
needed.  Foremost among them will be theoretical investigations, especially 
of magnetic dynamos in the envelopes and disks of Be stars. 
These can lead to requirements for their maintenance and predictions of 
their durability. 
\\

  We wish to thank Dr. Gerrie Peters and Chris Shrader for helpful 
conversations and for Chris's permission to use a modified figure in this paper.
We express our great appreciation to two conscientious referees, as well as
to our editor, Dr. Lida Oskinova, for a careful reading. Each of these 
professionala have provided comments that have led to improvements of this 
review.  R.L.O. was supported by the Brazilian agencies CNPq (Universal Grants 
459553/2014-3) and INCTA (CNPq/FAPESP)

\section{Citations}
\label{Section 3}

\end{document}